\documentclass[%
 reprint,
longbibliography,
 amsmath,amssymb,
 aps,
]{revtex4-1}

\usepackage{graphicx}
\usepackage{dcolumn}
\usepackage{bm}
\usepackage{subfigure}
\usepackage{caption}
\usepackage[justification=justified,singlelinecheck=false]{caption}
\usepackage{color}

\preprint{APS/123-QED}
\begin{document}
\title{Competing quantum phases of hard-core boson with tilted dipole-dipole interaction}

\author{Huan-Kuang Wu$^{1}$ and Wei-Lin Tu$^{2}$}

\affiliation{
$^{1}$\textit{Department of Physics, Condensed Matter Theory Center and Joint Quantum Institute, University of Maryland, College Park, MD 20742, USA}\\
$^{2}$\textit{Institute for Solid State Physics, University of Tokyo, Kashiwa, Chiba 277-8581, Japan}
}

\date{\today}

\begin{abstract}
Different quantum phases of hard-core boson induced by dipole-dipole interaction with varying angles of polarization are discussed in this work. 
We consider the  two most influential leading terms with anisotropy due to the tilted polarization of the on-site boson in the square lattice. 
To ensure the concreteness of this truncation, we compare our phase diagrams, obtained numerically from cluster mean-field theory (CMFT) and infinite projected entangled-pair state (iPEPS), with that of the long-range interacting model from quantum Monte Carlo.
Next, we focus on the case where the azimuthal angle is fixed to $\phi = \pi/4$. 
Using the mean-field analysis where the quantum spin operators are replaced by $c$-numbers, we aim to search for the underlying phases, especially the supersolid.
Our results show a competing scenario mainly between two ordered phases with different sizes of unit cell, where first-order transition takes place in between them.
With the help of CMFT and variational iPEPS, the phase boundaries predicted by the mean-field theory are determined more precisely.
Our discoveries elucidate the possible underlying supersolid phases which might be seen in the ultracold experiments with strongly dipolar atoms. 
Moreover, our results indicate that an effective triangular optical lattice can be realized by fine tuning the polarization of dipoles in a square lattice.
\end{abstract}

\pacs{Valid PACS appear here}
\maketitle

\section{\label{sec:level1}Introduction}

The ultracold atomic gases \cite{Bloch, Bloch2, Windpassinger, Tomza}, thanks to the advance of cooling techniques, have become one of the promising platforms to study various physical scenarios where quantum effect is emphasized. 
At such low temperatures, exotic quantum phases can appear due to the reduction of thermal fluctuation, including the supersolid (SS) phase \cite{Penrose, Boninsegni}, which is characterized by the coexistence of superfluidity and solidity.
Recently, experimental groups have successfully observed the signatures of SS in Bose-Einstein condensates made of erbium and dysprosium gases with large dipole moments \cite{Tanzi, Bottcher, Chomaz, Natale, Tanzi2, GuoS}. Their discoveries have had huge impact to the physics community.

From the theoretical side, the ultracold bosonic atoms, such as $^{87}\text{Rb}$, in an optical lattice can be viewed as bosonic systems with soft- or hard-core characteristics, depending on the strength of the on-site repulsion.
Under the hard-core limit, previous studies have shown that for a Hubbard-like Hamiltonian in a square lattice with only nearest-neighbor (nn) hopping, one will need more than the nn interaction to stabilize the SS phases \cite{Batrouni, Hebert, Ng, Dang}. 
A similar effect can be seen by including the next-nearest-neighbor (nnn) hopping term \cite{Dong, ChenS}, where peculiar SS phases appear under the frustration. 
The effect of combining nnn hopping and interaction has been discussed \cite{ChenSS} and our recent work has revealed that various superfluid (SF) and SS phases can be seen in such extended Bose-Hubbard (EBH) Hamiltonian \cite{Tuh}. 

\begin{figure}[t]
\centering
\includegraphics[width=0.48 \textwidth]{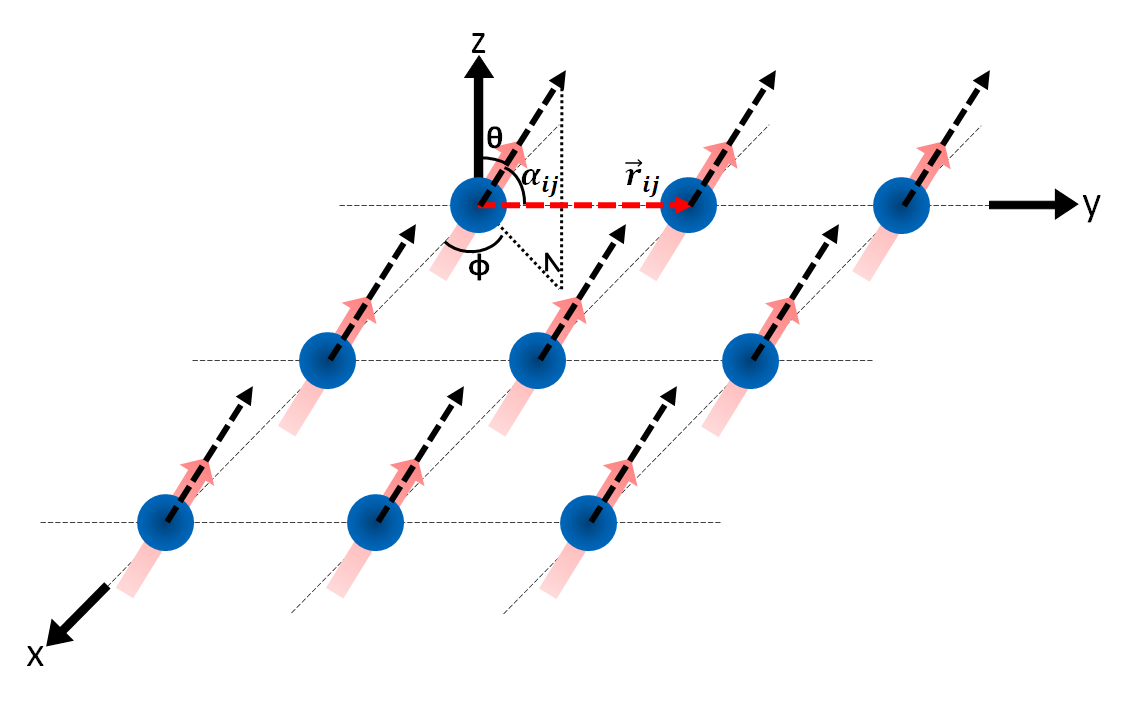}
\caption{\raggedright Schematic demonstration of dipolar interaction in a two dimensional square optical lattice. Blue dots denote the atomic sites and pink arrows represent the polarization. Notice that in this work we consider the hard-core limit so that on each site there will be at most one boson. Polarization vector (black dashed vector) is determined by polar ($\theta$) and azimuthal ($\phi$) angles. $\alpha_{ij}$ determines the anisotropic dipolar interaction in Eq. \ref{dipolar} and can be obtained by measuring the angle between polarization and the vector of relative positions (red dashed vector) between two sites.}
\label{Fig.1}
\end{figure} 

However, in the experiments mentioned above, the interaction between atoms is mediated by their intrinsic dipole moments \cite{Lahaye}. These dipole moments, as illustrated in Fig. \ref{Fig.1}, are polarized along the direction of external electric/magnetic field and introduce a long-range interaction of the following form:
\begin{equation}
\begin{aligned}
V_{ij}=\frac{V}{r_{ij}^3}(1-3\text{cos}^2\alpha_{ij}),
\end{aligned}
\label{dipolar}
\end{equation}
where $V$ is the dipole-dipole interaction strength and $r_{ij}=|\vec{r}_i-\vec{r}_j|$, is the relative distance between site $i$ and $j$. 
As shown in Fig. \ref{Fig.1}, $\alpha_{ij}$ represents the angle between the vector of polarization and $\vec{r}_{ij}$, the displacement vector from site $i$ to site $j$. 
Due to the external field, all the dipoles oriented in the same direction are parametrized by the polar ($\theta$) and azimuthal ($\phi$) angles. 
As can be seen from Eq. \ref{dipolar}, if the polarization is along $\hat{e}=(0,0,1)$, then $\alpha_{ij}=\pi/2$ between every $i$ and $j$ so that the repulsive interaction is isotropic and decreases in inverse cubic form with $r_{ij}$ throughout the optical lattice. Nonetheless, if we tilt the polarization to a non-zero polar angle, the interaction becomes anisotropic. 

The effect of dipole-dipole interaction on a square lattice has been discussed previously for isotropic \cite{Capogrosso, Ohgoe2, Yamamoto} and anisotropic \cite{Ohgoe, Zhangnjp,bandyopadhyay2019quantum} cases for the regular soft-core boson \cite{bandyopadhyay2019quantum} and in the hard-core limit \cite{Capogrosso, Ohgoe2, Yamamoto, Ohgoe, Zhangnjp}. In the absence of the lattice structure, there are also studies on the effect of the tilting angle of dipoles \cite{Macia, Macia2,bombin2017dipolar,cinti2019absence}. 
For the scenarios with anisotropic interaction on a square lattice, however, studies were mainly focused on the polarization within the x-z(y-z) plane under different polar angles. 
In Ref. \onlinecite{Zhangnjp}, the authors have mentioned the possible competition between two solid phases with intermediate azimuthal angle, which is the motivation of the present work. 
In fact, according to Eq. \ref{dipolar}, we can see that, as $\alpha_{ij}$ changes from $90^{\circ}$ to $0^{\circ}$, the dipolar interaction strength varies from $V/r^3_{ij}$ to $-2V/r^3_{ij}$, and is equal to zero at the transition angle $\text{cos}^{-1}(1/\sqrt{3})\approx 54.7^{\circ}$. 
If we choose $\phi$ to be $\pi/4$ and $35.3^{\circ}<\theta<\text{sin}^{-1}(\sqrt{2}\text{cos}(54.7^{\circ}))\approx 54.8^{\circ}$, we can ensure that the nn interaction is isotropically repulsive but the nnn interaction is attractive along the [1,1] direction while repulsive in the perpendicular direction. 
Under such circumstances, a diagonal stripe along [1,1] with a $3\times3$ unit cell \cite{Capogrosso, Yamamoto3} might be favored. 
Such an effect of anisotropic nnn terms leading to the translational-symmetry breaking into a larger unit cell has been demonstrated before in Ref. \onlinecite{Misumi, Huo}, which focus on the effect of nnn hopping terms, instead of the interaction terms. 
Moreover, although the dipole-dipole interaction is relatively long-range, due to the fact that the interaction strength decreases with distance in the inverse-cubic form, dominant short-range interactions are more influential in determining the resulting phases \cite{Chanpra}. 
This is reasonable because in the continuous limit, the long-range character is reflected by the integration $\int V(\textbf{r})d^2\textbf{r}$, for two-dimensional space. 
Due to the inverse-cubic form of $V(\textbf{r})$, the contribution from a large distance will vanish  and one can always find a cut-off range depending on the energy scale under consideration.
Therefore, we will focus on the effect of short-range interactions with realistic dipolar form written in Eq. \ref{dipolar}.

The rest of this paper is organized according to the following structure. In section II we will present our results. We first demonstrate our Hamiltonian and explain our numerical approaches in II.A. 
We then discuss the $\phi=0$ case and present the phase diagram in II.B.
Starting from II.C, we reveal our main results focusing on $\phi=\pi/4$. 
We will begin with a mean-field analysis to identify possible underlying phases, and a second-order perturbative approach to resolve the boundaries where solid order starts to melt down. 
We then perform the cluster mean-field theory to re-construct the phase diagrams. 
In order to attain the thermodynamic limit, we apply the variational iPEPS, which is even more precise than the simple update method. 
Finally, our discussion is included in section III.

\section{\label{sec:level1}Results}

\subsection{\label{sec:level2}Dipolar Bose-Hubbard Hamiltonian}

Our model describes a hard-core bosonic system in a square lattice with the following Hamiltonian:
\begin{equation}
\begin{aligned}
H=&-t\sum_{\langle i,j \rangle}(b^\dagger_{i}b_{j}+H.C.)\\
&+\sum_{\langle i,j\rangle}V_1 n_i n_j+\sum_{\langle \langle i,j \rangle \rangle}V_2 n_i n_j -\mu \sum_i n_i ,
\end{aligned}
\label{Hamiltonian}
\end{equation}
with
\begin{equation}
\begin{aligned}
&V_1=V(1-3\text{cos}^2\alpha_{ij})\\
&V_2=\frac{V(1-3\text{cos}^2\alpha_{ij})}{2\sqrt{2}},
\end{aligned}
\label{Interaction}
\end{equation}
where $b^\dagger_{i}$ and $b_{i}$ are the creation and annihilation operators of the hard-core boson, with the number operator to be $n_i=b^\dagger_{i}b_{i}$. $\langle i,j \rangle$ and $\langle \langle i,j \rangle \rangle$ denote the summation for the nn and nnn pairs, respectively. According to Eq. \ref{dipolar}, our inter-site interactions, $V_1$ and $V_2$, possess the forms in Eq. \ref{Interaction}. We do not include the further long-range tail of the dipolar interaction in this research. 

We adopt two numerical methods for solving our model and the first one is the cluster mean-field theory (CMFT) \cite{Gelfand, Hassan, Yamamoto2, Yamamoto3, McIntosh, Luhmann, Singh, Yamamoto, Moreno-Cardoner, Jurgensen, ChenS}.
The central spirit is quite similar despite the varying details or names used in different references.
We first divide our Hamiltonian into two parts. 
The first part, $H_C$, is within the chosen cluster, and the second part, $H_{\partial C}$, contains the terms connecting the bulk to the environment on the boundary of the cluster. 
$H_C$ possesses the exact form of the original Hamiltonian (Eq. \ref{Hamiltonian}) and the mean-field decoupling only takes place in $H_{\partial C}$:
\begin{equation}
\begin{aligned}
H_{\partial C}=&-t{\sum_{\langle i,j \rangle}}'(b^\dagger_{i}\langle b_{j}\rangle+H.C.)\\
&+{\sum_{\langle i,j\rangle}}'V_1 n_i \langle n_j\rangle +{\sum_{\langle \langle i,j \rangle \rangle}}'V_2 n_i \langle n_j\rangle,
\end{aligned}
\label{HamiltonianMF}
\end{equation}
where the prime indicates that this summation is between site $i$ on the boundary of the cluster and site $j$ connected to $i$ outside the cluster. 
Our effective Hamiltonian is then written as $H_{\text{eff}} = H_C + H_{\partial C}$. 
Next, we diagonalize the effective Hamiltonian and obtain the ground state to calculate the mean-field parameters, $\langle b_j \rangle$ and $\langle n_j \rangle$, for the next iteration. 
After several iterations, the mean-field parameters converge and our calculation reaches its self-consistent solution. 
Note that, the diagonalization only takes place within the chosen cluster.

The merit of CMFT is that, unlike the regular single-site mean-field calculation, CMFT can well cover the short-range correlation, since the solution within the cluster is exact. By gradually enlarging the cluster size, we can extrapolate the results to the thermodynamic limit and a more precise phase boundary can be obtained \cite{ChenS, Yamamoto2, Yamamoto3, Luhmann}. For solving the exact Hamiltonian within the cluster, we applied exact diagonalization (ED).

In addition to CMFT, we have also employed another numerical ansatz for comparison, which is the infinite projected entangled-pair state (iPEPS) \cite{Jordan}. 
iPEPS has been proven to be an effective ansatz for 2D quantum systems in a large scope \cite{Orus2}.
For this ansatz, we use rank-5 tensors to demonstrate the wave function on each site within the repeating unit cell. 
Each tensor is composed of four auxiliary legs with dimension $D$ and one physical leg with dimension $d=2$ here, reflecting the filled and empty states separately.
Another merit of iPEPS is that we can achieve the properties in the thermodynamic limit.
To attain the limit, we construct the environment tensors by applying the corner-transfer-matrix algorithm \cite{Corboz, Nishino, Orus}. 
In our previous work \cite{Tuh}, we have demonstrated that, for the extended Bose-Hubbard (EBH) model, simple-update iPEPS \cite{JiangS} is adequate for showing the correct structure of the phase diagram. 
Therefore, we will apply the same approach here.
However, later in section II.C.4, one will see that, for some regions where competing phases possess very close energies and it is hard for the simple-update iPEPS to distinguish them. 
In this case, we apply another optimization algorithm for the iPEPS, which is based on estimating the gradient of the variational energies to the tensor elements \cite{LiaoS, Hasik}.

\subsection{\label{sec:level2} $\phi$=0}

First of all, we revisit the previously studied case, where the polarization lies within the x-z(y-z) plane while tilting the polar angle. 
We will use this model to investigate the influence of removing the long-range tail.
For such polarization, the symmetry of nn interacting terms in x- and y-directions will be broken, while leaving the nnn interactions to be isotropic in our Hamiltonian:
\begin{equation}
\begin{aligned}
&V_{x}=V \\
&V_{y}=V(1-3\text{sin}^2\theta) \\
&V_{nnn}=\frac{V}{2\sqrt{2}}(1-\frac{3}{2}\text{sin}^2\theta).
\end{aligned}
\label{Interaction0phi}
\end{equation}
In Eq. \ref{Interaction0phi}, we have assumed that the polarization vector lies in the y-z plane.
We then try to reconstruct the phase diagram shown in Figure 2 of Ref. \onlinecite{Zhangnjp}, which focused on the half-filled doping. 
In fact, their phase diagram already implies the dominant influence from nn and nnn interactions, because the checkerboard (CB) and stripe solids break the translational symmetry within the 2 $\times$ 2 unit cell, which is also the unit cell for $V_{nn}$ and $V_{nnn}$ interaction. 
Therefore, we expect that our phase diagram will be similar to theirs.
\begin{figure*}[t]
\centering
\includegraphics[width=0.8 \textwidth]{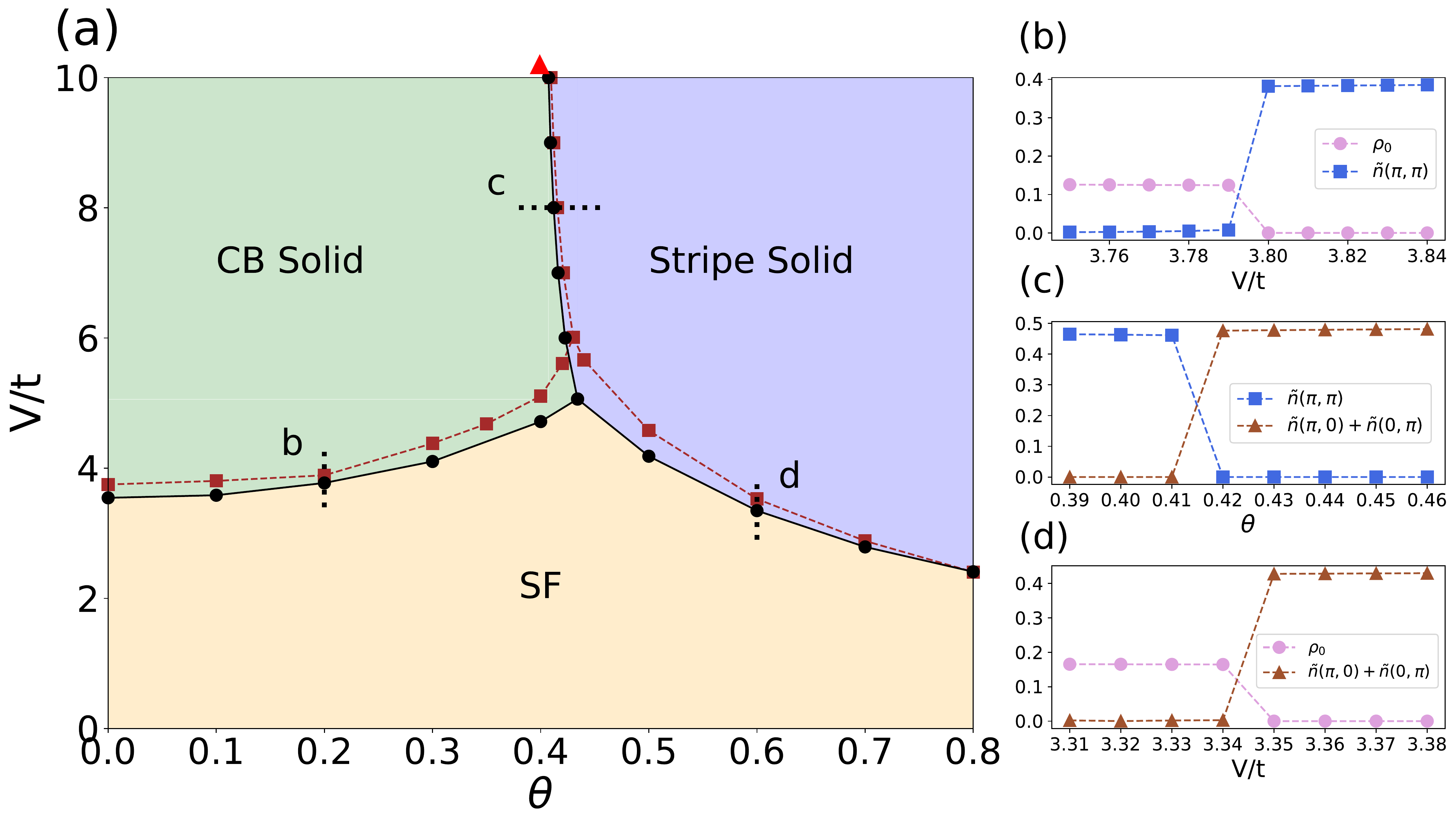}
\caption{\raggedright (a) Phase diagram for half-filled phases for $\phi = 0$ obtained from CMFT (black solid line) and iPEPS (brown dashed line). The red triangle on top of the diagram indicates the critical polar angle, $\theta_c$, between CB solid and stripe solid in the $V/t\rightarrow \infty$ limit. Order parameters by CMFT across three phase boundaries indicated by black dotted lines are plotted for (b) $\theta = 0.2$, (c) $V/t = 8$, and (d) $\theta = 0.6$.}
\label{Fig.2}
\end{figure*} 
Accordingly, we construct the phase diagram of half-filled phases using fixed-$D$ simple-update iPEPS (D=4) and CMFT with 4 $\times$ 4 clusters. 
Our order parameter for charge order is defined as
\begin{equation}
\begin{aligned}
\tilde{n}(\textbf{k})=\frac{1}{N_C}\sum_{i\in C}\langle n_{i} \rangle e^{i\textbf{k}\cdot \textbf{r}_i},
\end{aligned}
\label{DW}
\end{equation}
and for the condensate density,
\begin{equation}
\begin{aligned}
\rho_0= \frac{1}{N_C}\sum_{i\in C} | \langle b_{i} \rangle |^2,
\end{aligned}
\label{SF}
\end{equation}
where $C$ means the unit cell or cluster. 
$r_i$ is the coordinate of location for each site. 
The zero-momentum condensate density, $\rho_0$, indicates the superfluidity if a sharp peak exists.
$\tilde{n}(\textbf{k})$ reflects the structural order of bosons in the lattice. 
$\tilde{n}(\pi,\pi)$ represents the CB-like modulation and $\tilde{n}(\pi,0)$ ($\tilde{n}(0,\pi)$) stands for stripe-like order. 
Our states can be classified by calculating these two orders. 
If there is no structural order then the state corresponds to SF; on the other hand, we have a CB/stripe solid state once structural order exists and $\rho_0=0$.
Finally, the coexistence of $\rho_0$ and any structural orders $\tilde{n}(\textbf{k})$ with $\textbf{k}\neq 0$ implies the existence of an SS state.

Our result is shown in Fig. \ref{Fig.2}(a). The phase diagram for short-range dipolar interaction qualitatively agrees with the QMC result by C. Zhang et al. \cite{Zhangnjp}. The black solid lines are phase boundaries obtained from CMFT. 
At small $V/t$, the system is in SF phase, characterized by a non vanishing condensate density $\rho_0$. 
As $V/t$ is increased, the system goes through a first order phase transition and either enters a CB solid or a stripe solid phase, depending on the anisotropy of the nn interaction determined by $\theta$.
Since the interaction is weaker for nnn bonds, CB is more favored for smaller $\theta$. 
As $\theta$ increases, the anisotropy in nn interactions is enhanced. 
According to Eq. \ref{Interaction0phi}, when $\theta$ exceeds $\theta_c = \sin^{-1}\sqrt{2/(9+3\sqrt{2})}$, $V_y$ becomes smaller than $2V_{nnn}$ and the stripe solid state oriented in the y direction becomes more favorable in the classical limit.
$\theta_c$ is thus the critical polar angle in the limit $V/t\rightarrow \infty$, which is marked as the red triangle in Fig. \ref{Fig.2}(a). 
The phase transition between CB and stripe solid is first-order because of the breaking of different symmetries. 
Along this phase boundary, as $V/t$ decreases, the kinetic energy becomes more important. 
From the perturbation points of view, the self-energy gain from virtual hopping process is $-t^2/2V_x-t^2/(4V_x-2V_y)$ for CB solid and $-t^2/2V_x$ for stripe solid. 
This causes CB states to be more favorable, making the line lean to the right slightly as approaching the triple point. 
Moreover, the phase boundaries by iPEPS (brown dashed lines in Fig. \ref{Fig.2}(a)) agree with those from CMFT, indicating that the underlying physics can be well represented by the finite-size calculation.
Finally, it is important to note that our calculations capture the correct phases revealed by QMC upon the full long-range dipolar model \cite{Zhangnjp}.

We next demonstrate the transitions across three phase boundaries along the cuts indicated by the black dotted lines in Fig. \ref{Fig.2}(a). 
Fig. \ref{Fig.2}(b) shows the condensate density and CB structural order parameters along the vertical cut $\theta = 0.2$. 
For smaller $V/t$, there is finite $\rho_0$ and no CB structural factor, which represents an SF phase. 
At around $V/t \sim 3.8$, $\rho_0$ disappears with the simultaneous onset of $\tilde{n}(\pi,\pi)$, suggesting a first order transition to CB solid phase. 
Fig. \ref{Fig.2}(c) shows the CB and stripe structural factor along the horizontal cut at $V/t = 8$, where the two solid phases compete. 
Again, the order parameters indicate first-order features with an abrupt change from $\tilde{n}(\pi,\pi) \neq 0$ (CB solid) to $\tilde{n}(0,\pi)+\tilde{n}(\pi,0) \neq 0$ (stripe solid) at around $\theta = 0.415$. 
At last, Fig. \ref{Fig.2}(d) is the vertical cut along $\theta = 0.6$. 
Similar to \ref{Fig.2}(b), the system starts from the SF phase with non-zero $\rho_0$ and vanishing $\tilde{n}(\pi,\pi)$, but it transits to stripe solid as $\tilde{n}(0,\pi)+\tilde{n}(\pi,0)$ becomes finite at stronger interaction. 
Note that, the order parameters showing discontinuous jumps at the transition points in Fig. \ref{Fig.2}(b), (c), and (d) is a signature of the first-order phase transitions.
The facts that all of our phase boundaries are first-order and that no SS phase appears are in agreement with Ref. \onlinecite{Zhangnjp}.

The above results show that our short-range dipolar Hamiltonian is able to reflect the same solid phases and behavior of the full dipolar model if the repeating unit cell is 2 $\times$ 2, despite some quantitative differences.
This accords with our expectation. 
The difference is that with the current Hamiltonian, we are not able to deal with the small plateaus, named after the Devil's staircase, in the $\mu-t$ phase diagram \cite{Capogrosso, Ohgoe}.
These plateaus break higher symmetries under the long-range interaction, which is excluded here.
However, in this work we strive to investigate the dominant phases and their properties.
Therefore, it is reasonable to remove the longer-range tail in the current scope.
After we present our main results in the next section, we will come back and conjecture upon how our results would change with the presence of the long-range interaction in the discussion section.

We conclude the discussion of short-range dipolar interaction with dipoles lying in the x-z(y-z) plane. Next, we will investigate its effect when the azimuthal angle is set equal to $\pi/4$, where a more fruitful phase diagram appears.

\subsection{\label{sec:level2} $\phi=\pi/4$}

In contrast to the previous case, where polarization lies in the x-z(y-z) plane, by choosing $\phi=\pi/4$ we will have isotropic nn interacting terms, while nnn terms become anisotropic:
\begin{equation}
\begin{aligned}
H=&-t\sum_{\langle i,j \rangle}(b^\dagger_{i}b_{j}+H.C.) +V_{nn}\sum_{\langle i,j\rangle} n_i n_j\\
&+V_{[1,1]}\sum_{\langle \langle i,j \rangle \rangle_{[1,1]}} n_i n_j +V_{[1,-1]}\sum_{\langle \langle i,j \rangle \rangle_{[1,-1]}} n_i n_j \\
&-\mu \sum_i n_i ,
\end{aligned}
\label{Hamiltonian1/4phi}
\end{equation}
where
\begin{equation}
\begin{aligned}
&V_{nn}=V(1-\frac{3}{2}\text{sin}^2\theta)\\
&V_{[1,1]}=\frac{V(1-3\text{sin}^2\theta)}{2\sqrt{2}} \\
&V_{[1,-1]}=\frac{V}{2\sqrt{2}}.
\end{aligned}
\label{Interaction1/4phi}
\end{equation}
[1,1]/[1,-1] denotes the direction of the nnn interaction. 
Since we have set $\phi=\pi/4$ here, the interaction terms are expressed with one variable, the polar angle $\theta$ (Eq. \ref{Interaction1/4phi}).
We will examine the effect of altering $\theta$ in this model. 
Note that, according to Eq. \ref{Interaction1/4phi}, $V_{[1,-1]}$ is independent of $\theta$ while $V_{[1,1]}$ will alter with $\theta$ and change from repulsive to attractive interaction. 
We will then focus on the region of intermediate $\theta$. 

\subsubsection{\label{sec:level3} Mean-field analysis}

\begin{figure}[t]
\centering
\includegraphics[width=0.48 \textwidth]{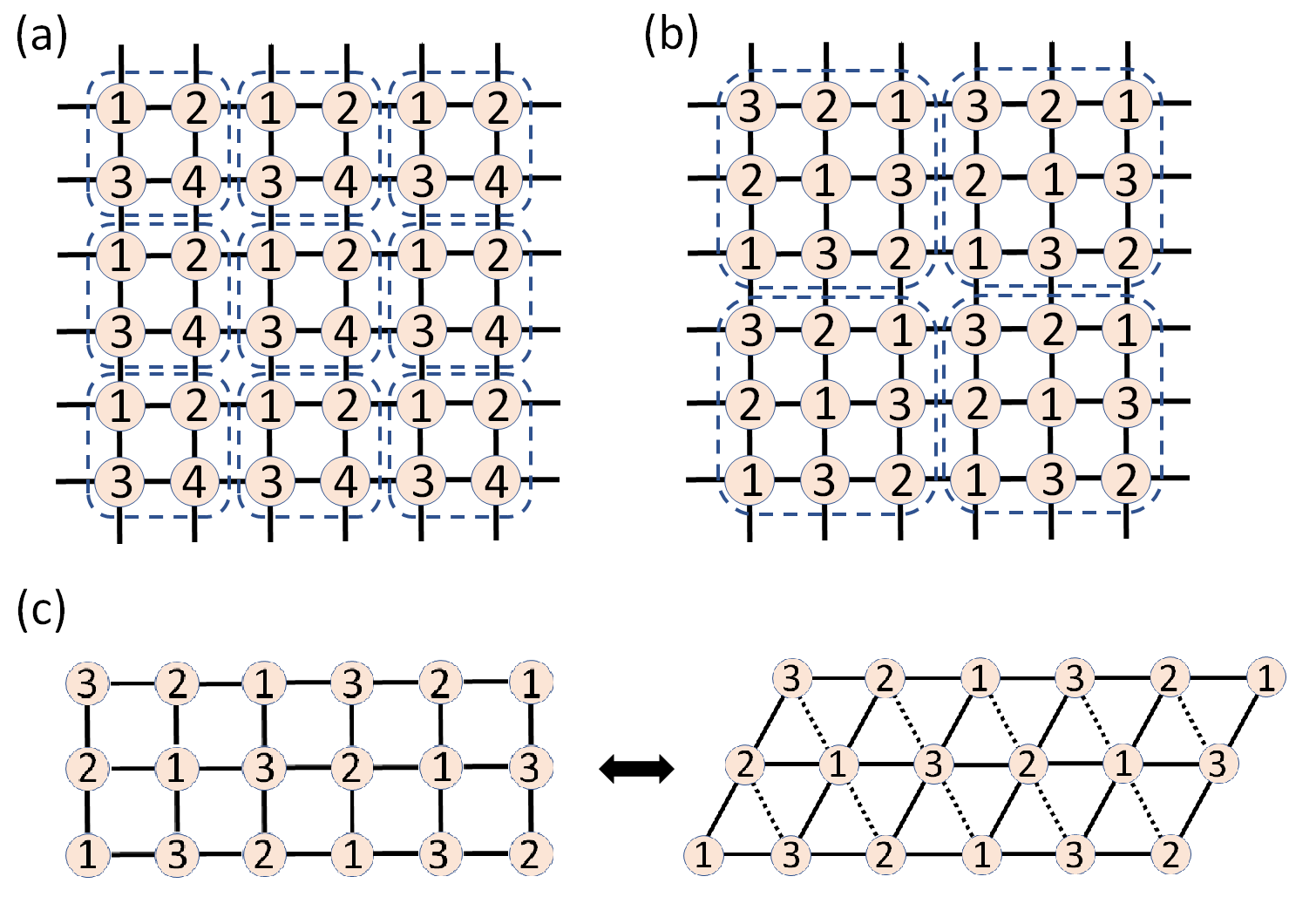}
\caption{\raggedright Schematic demonstration of sublattice structures for (a) 2 $\times$ 2 and (b) 3 $\times$ 3 unit cell (dashed boxes) that we adopt for the mean-field analysis. For 2 $\times$ 2 unit cell we divide lattice sites into four different kinds, which guarantee to include all possible underlying structures. As for the 3 $\times$ 3 unit cell, we have three kinds of sublattices. (c) The equivalence of an eighteen-site square lattice to the effective triangular lattice with the same number of sites. Dotted lines indicate the virtual bonds where hopping is not allowed for bosons.}
\label{Fig.3}
\end{figure} 

\begin{figure*}[t]
\centering
\includegraphics[width=1.0 \textwidth]{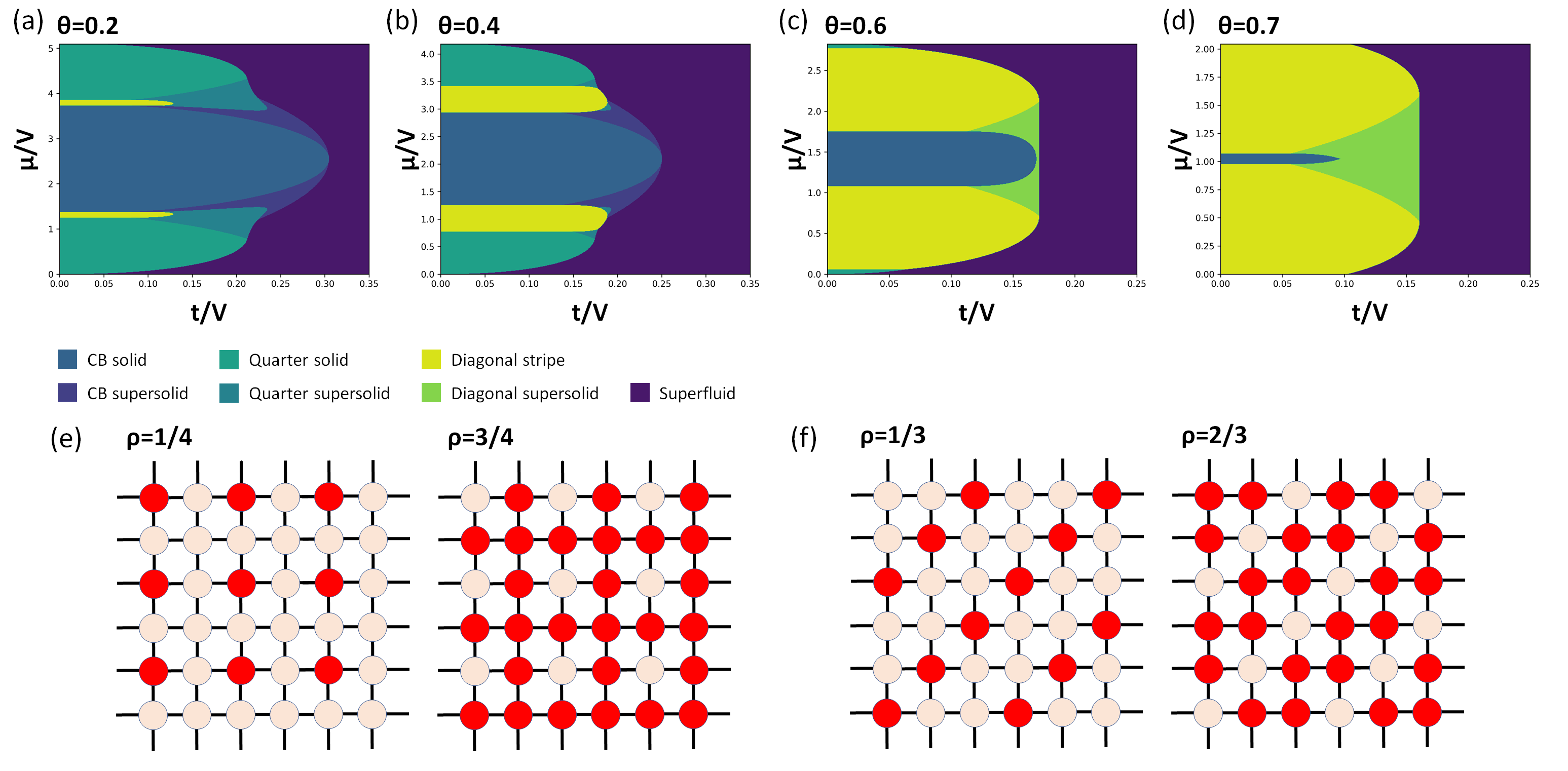}
\caption{\raggedright Mean-field phase diagrams with polar angles (a) $\theta=0.2$, (b) $\theta=0.4$, (c) $\theta=0.6$, and (d) $\theta=0.7$. Figure legends are placed in the middle panel. Notice that in fact we have two quarter($\rho=1/4$ and $3/4$) and diagonal stripe($\rho=1/3$ and $2/3$) solids/SSs, but we label them with identical colors to demonstrate the particle-hole symmetry. (e) and (f) show the structures of these two quarter and diagonal stripe solids. Note that in (e) the configurations stay the same if we shift one lattice constant for all the occupied($\rho=1/4$)/empty($\rho=3/4$) sites on a row(column) along the horizontal(vertical) direction. Sites with darker color are the occupied sites and the others are for empty sites.}
\label{Fig.4}
\end{figure*} 

To properly discuss possible underlying phases, it would be better to start from a mean-field analysis. 
The hard-core EBH model is known for its connection with the spin-1/2 XXZ model under the mapping of $b_i^\dagger \to \hat{S}_i^{+}$, $b_i \to \hat{S}_i^{-}$, and $n_i \to \hat{S}_i^{z}+1/2$ \cite{Matsubara, Lieb}. 
Therefore, our Hamiltonian can be rewritten in the following form:
\begin{equation}
\begin{aligned}
H=&-2t\sum_{\langle i,j \rangle}(\hat{S}_i^{x}\hat{S}_j^{x}+\hat{S}_i^{y}\hat{S}_j^{y})+V_{nn}\sum_{\langle i,j\rangle}\hat{S}_i^{z}\hat{S}_j^{z}\\
&+V_{[1,1]}{\sum_{\langle \langle i,j \rangle \rangle_{[1,1]}}}\hat{S}_i^{z}\hat{S}_j^{z}+V_{[1,-1]}{\sum_{\langle \langle i,j \rangle \rangle_{[1,-1]}}} \hat{S}_i^{z}\hat{S}_j^{z} \\
&-(\mu-2V_{nn}-V_{[1,1]}-V_{[1,-1]}) \sum_i \hat{S}_i^{z}.
\end{aligned}
\label{Spin_Hamiltonian}
\end{equation}
Now the Hamiltonian is represented with the pseudospin operator $\hat{\textbf{S}}_i=(\hat{S}_i^{x},\hat{S}_i^{y},\hat{S}_i^{z})$, which satisfies the commutation relation:
\begin{equation}
\begin{aligned}
\lbrack \hat{S}_i^{\mu},\hat{S}_j^{\nu} \rbrack=i\epsilon_{\mu\nu\lambda}\hat{S}_i^{\lambda}\delta_{ij}.
\end{aligned}
\label{Commutation}
\end{equation}

In contrast to the ordinary spin-1/2 XXZ model, Eq. \ref{Spin_Hamiltonian} contains anisotropic interaction terms.
For such a spin model, the last term in Eq. \ref{Spin_Hamiltonian} can be viewed as a Zeeman term with an effective magnetic field $h=\mu-2V_{nn}-V_{[1,1]}-V_{[1,-1]}$. 
Related physical observables of the hard-core boson can be evaluated with $\langle n_i \rangle = \langle \hat{S}^z_i \rangle+1/2$ and $\langle b_i \rangle = \langle \hat{S}^{-}_i \rangle$. 
Therefore, after the mapping, we can adopt the treatment of a spin model and then interpret our results back to the hard-core bosonic side. 

At zero temperature, we can apply a mean-field treatment to the spin model by replacing the pseudospins with classical spin vectors with magnitude $S$ equal to 1/2 \cite{Yamamoto3}:
\begin{equation}
\begin{aligned}
\hat{\textbf{S}}_i \to \textbf{S}^{cl}_i=S(\text{cos}\varphi_i\text{sin}\vartheta_i,\text{sin}\varphi_i\text{sin}\vartheta_i,\text{cos}\vartheta_i).
\end{aligned}
\label{Classical}
\end{equation}
After such transformation, we can express the mean-field energy, $E^{\text{MF}}$, of Eq. \ref{Spin_Hamiltonian} in terms of different kinds of on-site pseudospin directions ($\vartheta$,$\varphi$). 
Note that, to properly reproduce desired states with distinct structural order, we need to select a correct sublattice structure. 
Due to the rotational symmetry of local spin in the x-y plane, we can take $\varphi=0$ for all sites without loss of generality. 
We first consider four kinds of sublattice sites to investigate states which break the translational symmetry within the 2 $\times$ 2 unit cell. The repeated unit cells form a square lattice as illustrated in Fig. \ref{Fig.3}(a). By substituting Eq. \ref{Classical} into Eq. \ref{Spin_Hamiltonian}, we can calculate the total energy within a unit cell.
We then take an average by dividing the total energy with the number of sites within the unit cell and obtain the mean-field energy, $E^{\text{MF}}_4$, for a single site:
\begin{equation}
\begin{aligned}
E^{\text{MF}}_{4}=&-\frac{1}{4}t(\text{sin}\vartheta_1\text{sin}\vartheta_2+\text{sin}\vartheta_1\text{sin}\vartheta_3\\
&+\text{sin}\vartheta_2\text{sin}\vartheta_4+\text{sin}\vartheta_3\text{sin}\vartheta_4)\\
&+\frac{1}{8}V_{nn}(\text{cos}\vartheta_1\text{cos}\vartheta_2+\text{cos}\vartheta_1\text{cos}\vartheta_3\\
&+\text{cos}\vartheta_2\text{cos}\vartheta_4+\text{cos}\vartheta_3\text{cos}\vartheta_4)\\
&+\frac{1}{8}V_{[1,1]}(\text{cos}\vartheta_1\text{cos}\vartheta_4+\text{cos}\vartheta_2\text{cos}\vartheta_3)\\
&+\frac{1}{8}V_{[1,-1]}(\text{cos}\vartheta_1\text{cos}\vartheta_4+\text{cos}\vartheta_2\text{cos}\vartheta_3)\\
&-\frac{1}{8}h(\text{cos}\vartheta_1+\text{cos}\vartheta_2+\text{cos}\vartheta_3+\text{cos}\vartheta_4),
\end{aligned}
\label{E4}
\end{equation}
where the subscripts of $\vartheta$'s are the site indices within the unit cell. 
Note that, we have already included the condition that $S=1/2$ into our energy. 
In Ref. \onlinecite{Yamamoto3}, the authors have also considered the 2 $\times$ 2 unit cell for only two sublattices to describe the CB or stripe solid. 
But if we adopt four sublattices, then all possible underlying structures, such as CB, stripe, or quarter($\rho=1/4$ or $3/4$) solids, can be accounted for. 

Because the dipolar interaction is anisotropic here, although our Hamiltonian contains no terms more than the next nearest neighbor, the stable states can still break the translational symmetries of larger unit cell. K. Misumi et al. \cite{Misumi} have shown that such anisotropy can lead to the symmetry breaking within the 3 $\times$ 3 unit cell, which is in fact effectively equivalent to the triangular lattice \cite{Zhangprb}. 
For our model, we expect that such equivalence would appear as the effect of nnn interacting term in one direction ($V_{[1,1]}$) becomes more dominant than the other ($V_{[1,-1]}$), while it varies from repulsion to attraction along with the polar angle $\theta$. 
To address such states, we also consider the case of three sublattices , whose lattice structure is shown in Fig. \ref{Fig.3}(b). In this case, the energy per site is given by
\begin{equation}
\begin{aligned}
E^{\text{MF}}_{3}&=\\
&-\frac{1}{3}t(\text{sin}\vartheta_1\text{sin}\vartheta_2+\text{sin}\vartheta_1\text{sin}\vartheta_3+\text{sin}\vartheta_2\text{sin}\vartheta_3)\\
&+\frac{1}{6}V_{nn}(\text{cos}\vartheta_1\text{cos}\vartheta_2+\text{cos}\vartheta_1\text{cos}\vartheta_3+\text{cos}\vartheta_2\text{cos}\vartheta_3)\\
&+\frac{1}{12}V_{[1,1]}(\text{cos}^2\vartheta_1+\text{cos}^2\vartheta_2+\text{cos}^2\vartheta_3)\\
&+\frac{1}{12}V_{[1,-1]}(\text{cos}\vartheta_1\text{cos}\vartheta_2+\text{cos}\vartheta_1\text{cos}\vartheta_3+\text{cos}\vartheta_2\text{cos}\vartheta_3)\\
&-\frac{1}{6}h(\text{cos}\vartheta_1+\text{cos}\vartheta_2+\text{cos}\vartheta_3).
\end{aligned}
\label{E3}
\end{equation}
Such a three-sublattice structure can be viewed as an effective triangular lattice, shown in Fig. \ref{Fig.3}(c). 
The diagonal bonds are virtual since our Hamiltonian does not include the nnn hopping. 
However, the nnn interactions, $V_{[1,1]}$ and $V_{[1,-1]}$, can still be present between diagonal bosons.
Other kinds of symmetry breaking for larger unit cells will require the inclusion of longer-range interaction.
Note that, in this work we do not consider the possibility of incommensurately ordered phases.
We will then use the mean-field energies in Eq. \ref{E4} and \ref{E3} to construct the phase diagrams under different polar angles. Since these two energies have different minimum values, we need to search for the ground-state energies for both cases and then compare the values to decide which one should be the true ground state. 

In Fig. \ref{Fig.4}(a) to (d), we plot the mean-field phase diagrams for different polar angles. 
For $\theta = 0.2$ and $0.4$, the phase diagrams are mainly composed of states with 2 $\times$ 2 unit cell, resembling the results with isotropic interaction \cite{Capogrosso}. 
When we further increase the polar angle to $\theta=0.6$, the phase of diagonal stripe solid becomes predominant.
Moreover, between the lobes of CB and diagonal stripe solids, diagonal SS phase appears. 
Finally, for $\theta=0.7$, diagonal stripe phases almost occupy the whole phase diagram and the CB phase only appears in a very narrow region. 
This kind of evolving process for the phase diagrams is consistent with our previous discussion. 
The main effect of increasing $\theta$ lies in the enhancement of the anisotropy for the interaction and therefore changes the nnn interaction in the $[1,1]$ direction from repulsive to attractive one. 
This anisotropy leads to the formation of the diagonal stripe phases, which possess a unit cell with larger size than the range of interaction. 
If we further tilt the polar angle, then after $\theta \approx 0.956$, even nn interaction becomes attractive.
As a result, the phase diagram becomes trivial for $\mu>0$, where all sites are fully occupied.

Notice that, unlike the case when polarization is within the x-z(y-z) plane, we cannot find any traditional stripe phase. 
This can be demonstrated by comparing the mean-field energies of CB and stripe solids, setting $\text{cos}\vartheta_1=\text{cos}\vartheta_4=-\text{cos}\vartheta_2=-\text{cos}\vartheta_3=1$ for CB and $\text{cos}\vartheta_1=\text{cos}\vartheta_2=-\text{cos}\vartheta_3=-\text{cos}\vartheta_4=1$ for stripe:
\begin{equation}
\begin{aligned}
E^{\text{MF(CB)}}_{4}=-\frac{1}{2}V_{nn}+\frac{1}{4}(V_{[1,1]}+V_{[1,-1]}),
\end{aligned}
\label{E4CB}
\end{equation}
and
\begin{equation}
\begin{aligned}
E^{\text{MF(stripe)}}_{4}=-\frac{1}{4}(V_{[1,1]}+V_{[1,-1]}).
\end{aligned}
\label{E4st}
\end{equation}
Since the condition for forming the stripe solid is that $E^{\text{MF(CB)}}_{4}>E^{\text{MF(stripe)}}_{4}$, therefore we obtain $V_{nn}<V_{[1,1]}+V_{[1,-1]}$. Then according to Eq. \ref{Interaction1/4phi}, the condition becomes $\theta \gtrsim 0.955$, where the states that break 2 $\times$ 2 translational symmetry are no longer prominent. 
As a result, no stripe solid/SS can be formed. 

In fact, by analyzing the mean-field energy, the location of the phase boundaries and whether they are first- or second-order transitions can be determined analytically. This has been presented in the case of a 2 $\times$ 2 unit cell in Ref. \cite{Yamamoto3}. 
Here, we will provide a similar analysis of the complementary case for the 3 $\times$ 3 supercell. 
This will be done by examining the mean-field energy $E_3^{MF}$ in Eq. \ref{E3}. Since, under the particle-hole transformation ($\vartheta_i \rightarrow \vartheta_i+\pi, i = 1,2,3$), all terms remain the same except for the Zeeman field $h\rightarrow -h$, we only have to consider the lower-half of the phase diagram where $h$ is negative. In this case, the three states to be considered are SF ($\vartheta_1 = \vartheta_2=\vartheta_3$), $1/3$ diagonal stripe ($\vartheta_1=0$, $\vartheta_2=\vartheta_3=\pi$) and the diagonal SS (all the others). 

The phases shown in Fig. \ref{Fig.4}(a)-(d) represent the global minima of energy in the phase space. 
For the phase transitions between adjoint phases, there are two possibilities, first- and second-order phase transitions.
For the first-order transition, there can be different points in the phase space with competitively low energies at the same time, while possessing different order parameters. 
In this case, we have a discontinuous transition, which is the feature for first-order transition.
Another possibility is when the ground state point becomes a saddle point, which is no longer a local minimum. 
In this case, the transition is continuous and of second order. 
Such phase boundaries are characterized while the determinant of the Hessian matrix, $\mathbf{H}_{i,j}(\vartheta_1,\vartheta_2,\vartheta_3) = \partial^2 E^{MF}_3/\partial \vartheta_i\partial\vartheta_j$, is equal to zero. 
We first look at the latter case for the diagonal stripe solid and SF. 
For diagonal stripe solid phase, the corresponding pseudospin coordinate, $(\vartheta_1,\vartheta_2,\vartheta_3)$, is located at $(0,\pi,\pi)$. 
Note that, one can also choose the other two configurations ($(\pi,0,\pi)$ and $(\pi,\pi,0)$) while leaving the following analysis unchanged.
On the other hand, for SF we have $(\vartheta_1,\vartheta_2,\vartheta_3)=(\psi,\psi,\psi)$, where $\psi = \cos^{-1}(h/4t+2V_{nn}+V_{[1,1]}+V_{[1,-1]})$.
This is the point that minimizes $E_3^{MF}$ along $\vartheta_1=\vartheta_2=\vartheta_3$. Plugging in these two conditions separately, we found that the position where $det(\mathbf{H})=0$ for diagonal stripe solid satisfies:
\begin{align}
    8t^2 + (2t + V_{[1,1]} + h)(2V_{nn}-V_{[1,1]}+V_{[1,-1]}+h) = 0,
    \label{mf_boundary}
\end{align}
and for the SF:
\begin{align}
    &h^4(4t+2V_{nn}-2V_{[1,1]}+V_{[1,-1]})\nonumber\\
    &+2h^2(4t + 2V_{nn}+V_{[1,1]}+V_{[1,-1]})^2\nonumber\\
    &\qquad\qquad\times(2t-2V_{nn}-V_{[1,-1]}+2V_{[1,1]})\nonumber\\
    &-(4t + 2V_{nn}+V_{[1,1]}+V_{[1,-1]})^4\nonumber\\
    &\qquad\qquad\times(8t-2V_{nn}+2V_{[1,1]}-V_{[1,-1]}) = 0.
\end{align}
Note that, although these two conditions will appear as two curves in the $t-\mu$ figure, they merely describe points where a particular phase is no longer a local minimum.
Therefore, if the phases (diagonal stripe solid or SF) are not the global minima in the first place, the curves will not show up as phase boundaries in the phase diagram. 
On the other hand, if we see a coincident segment of these curves and the phase boundaries, we could be sure that such phase boundaries are of second order.

\begin{figure}[t]
\centering
\includegraphics[width=0.48 \textwidth]{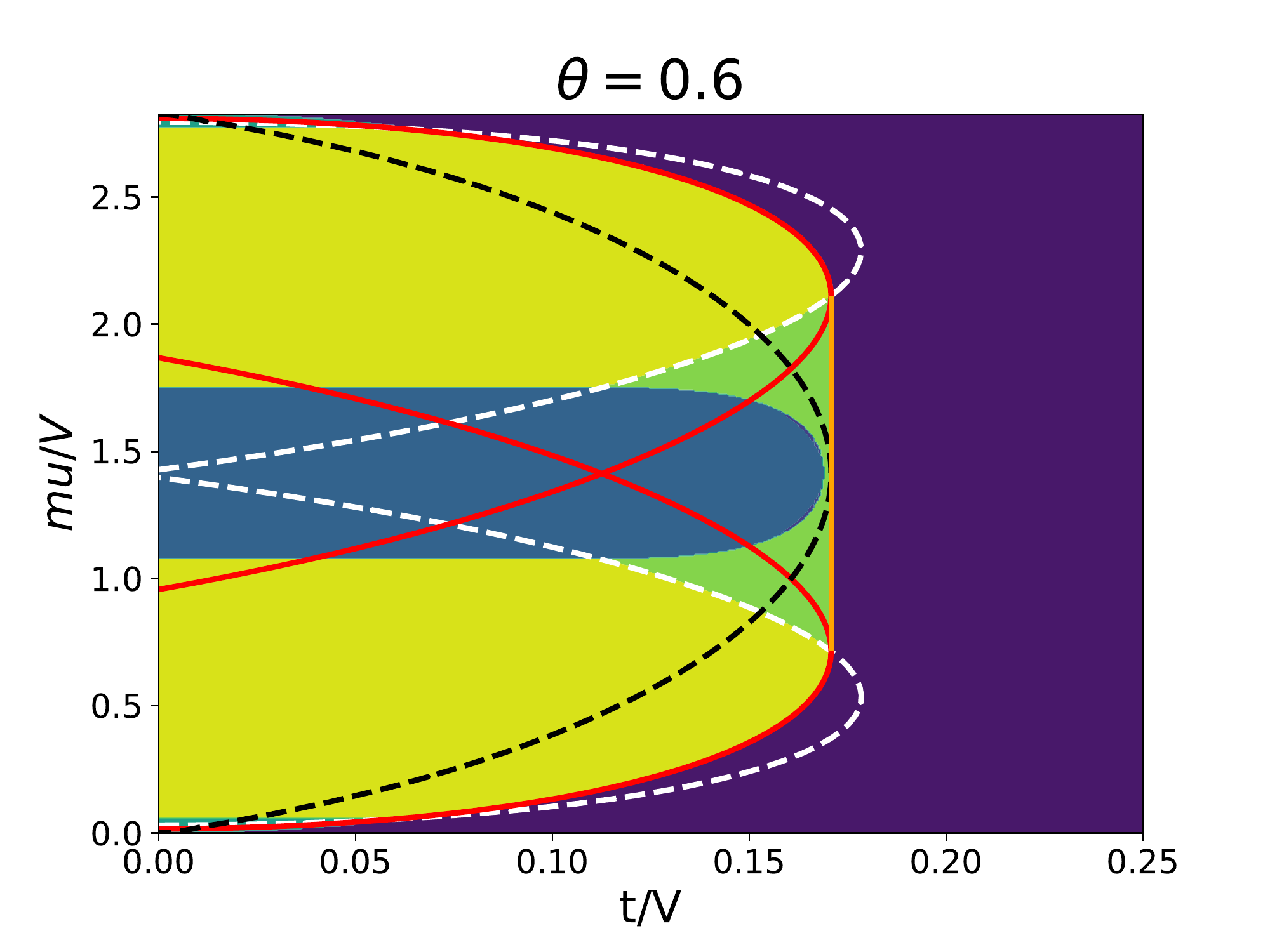}
\caption{\raggedright Analytic phase boundaries for tilting angle $\theta = 0.6$. For comparison, we adopt the corresponding phase diagram in Fig. \ref{Fig.4}(c) as the background. Boundaries indicated by solid(dashed) curves are of first(second) order. The white dashed lines corresponds to $det(\mathbf{H(0,\pi,\pi)})=0$ while the black dashed curve corresponds to $det(\mathbf{H(\psi,\psi,\psi)})=0$.
Red(Orange) solid curves(line) represent the points of energy crossover between diagonal stripe solid(SS) and SF, where the enclosed area on the left-hand side always breaks translational symmetry.}
\label{Fig.5}
\end{figure} 

In Fig. \ref{Fig.5}, we adopt the case of polar angle $\theta = 0.6$ as an example to demonstrate different analytical boundaries. 
Except for the first-order boundaries surrounding the checkerboard solid phases, all the other phase boundaries can be accounted for within our analytical evaluation. 
First, the two curves for $det(\mathbf{H})=0$ mentioned above are presented as dashed lines.
The white dashed lines enclose two regions where diagonal stripe solid is an energetic local minimum; on the other hand, outside the region enclosed by the black dashed line, the SF state is a local minimum of the mean-field energy. 
The white curves coincide with the phase boundaries between diagonal stripe solid and SS, suggesting that the transition between these two phases is second-order.

Since the diagonal stripe solid and SF states break different symmetries, the phase boundaries between them must be of first order, and thus do not belong to continuous dashed boundaries. In this case, we would need to compare the energies of the two states to determine the boundaries. Their analytic form can be derived by simply setting $E_3^{MF}(0,\pi,\pi) = E_3^{MF}(\psi,\psi,\psi)$, which results in the formula:
\begin{align}
    12t-2V_{nn}&+3V_{[1,1]}-V_{[1,-1]} + 2h \nonumber\\ &+\frac{3h^2}{4t+2V_{nn}+V_{[1,1]}+V_{[1,-1]}}=0.
\end{align}
This corresponds to the red solid lines in Fig. \ref{Fig.5}. Again, this curve is not necessarily the final phase boundary but we can identify that the transition is first-order if there is an overlapping with the phase boundary.

Finally, the boundary between diagonal SS and SF is less obvious due to the fact that the diagonal SS does not have a strong constraint for its degrees of freedom as in the other two phases. However, we notice that the tips of the two red solid curves and the black dashed curve are all located at a vertical straight line, which is
\begin{align}
    t = \frac{2V_{nn}+V_{[1,-1]}-2V_{[1,1]}}{8}.
    \label{Eq.19}
\end{align}
Plugging this back into Eq. (\ref{E3}), the mean field energy becomes
\begin{align}
    &E_3^{MF} \nonumber\\
    &= \frac{2V_{nn}+V_{[1,-1]}-V_{[1,1]}}{24}[(\sin \vartheta_1-\sin \vartheta_2)^2\nonumber\\
    &\qquad\qquad\qquad+(\sin \vartheta_2-\sin \vartheta_3)^2+(\sin \vartheta_3-\sin \vartheta_1)^2]\nonumber\\
    &+\frac{2V_{nn}+V_{[1,-1]}}{24}(\sum_i\cos \vartheta_i-\frac{2h}{2V_{nn}+V_{[1,-1]}})^2\nonumber\\
    &+ \frac{2V_{[1,1]}-2V_{nn}-V_{[1,-1]}}{8}-\frac{h^2}{6(V_{[1,-1]}+2V_{nn})}.
\end{align}
According to the above form, there are two global minima in the phase space where the first two terms vanishes. One is an SF state at $\vartheta_1 = \vartheta_2 = \vartheta_3 =\cos^{-1}(2h/(6V_{nn}+3V_{[1,-1]}))$ and the other is a diagonal SS on $\vartheta_1 = \pi-\vartheta_2 = \pi-\vartheta_3$ with $\vartheta_1 = \cos^{-1}(-2h/(2V_{nn}+V_{[1,-1]}))$.  As a result, the straight line of Eq. (\ref{Eq.19}), denoted by the orange color in Fig. \ref{Fig.5}, indicates the first-order boundary between the diagonal SS and SF.
Note that, the point of intersection between the orange line and the black dashed curve is a highly symmetrical point, where the breaking of translational symmetry can take place continuously.
Similar first-order transition between diagonal SS and SF is also observed in Refs. \onlinecite{Zhangprb, Yamamoto2} and it can not be directly inferred from the symmetry argument.
At last, since the states with a 3 $\times$ 3 unit cell compete with those with a 2 $\times$ 2 unit cell, part of the phase boundaries is replaced by the one caused from the phase competition. Note that, since these competing phases always break different symmetries, the phase boundaries are of first order.

In sum, our mean-field analysis reveals that the physical scenario we discuss here can be interpreted as the competition between states belonging to square and effectively triangular lattices. This is largely different from the previous case when polarization is within the plane of the principal axis. Note that, since quantum fluctuations are neglected in the mean-field analysis, it only provides a qualitative understanding. This is the reason for using CMFT and iPEPS, which can provide more accurate phase diagrams.

\subsubsection{\label{sec:level3} Defect condensation}
\begin{figure}[t]
\centering
\includegraphics[width=0.48 \textwidth]{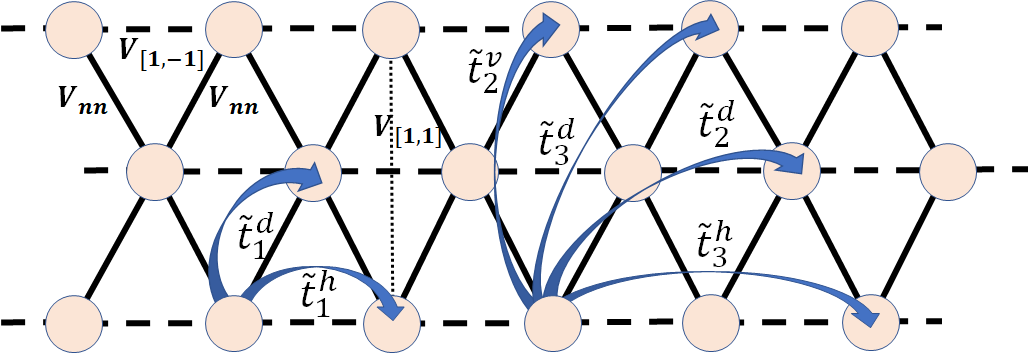}
\caption{\raggedright Schematic demonstration of interaction and hopping terms after mapping to the effective triangular lattice. $V_{nn}$ lies on the diagonal bonds and $V_{[1,-1]}$/$V_{[1,1]}$ lies on the virtual horizontal(dashed)/vertical(dotted) bonds. Hopping terms for the second-order perturbation theory are also plotted for nearest, next-nearest, and next-next-nearest neighbor, which are generated by two-step processes.}
\label{Fig.6}
\end{figure} 
Our mean-field analysis has revealed the existence of solids and SSs, including the diagonal SS state from the three-sublattice scenario. Therefore, it is worth investigating more of the effective triangular lattice along with the related phases, and the perturbation theory is helpful for showing the phase transition from solid to SS \cite{Zhangprb}. Because the SS is formed by doping the commensurate solid, called the ``defect-condensation" \cite{ChenSS}, we can think of doping as adding defects into the background composed of a perfect solid. By studying the energy of the defect we can determine the transition points where having defects within a lattice is more stable, leading to the formation of SS. For that purpose, first we need to write down the effective model for the defects. 

Now, with focus on the effective triangular lattice, the original interaction terms for the square lattice are mapped into the form shown in Fig. \ref{Fig.6}. For the  two nearest sites, interactions exist along the diagonal ($V_{nn}$) and horizontal ($V_{[1,1]}$) directions, while $V_{[1,-1]}$ corresponds to the interaction across the diamond. Due to the particle-hole symmetry for the hard-core boson, we only analyze the scenario when bosons are doped as defects into the 1/3 diagonal stripe. In such a scenario, defects lie on a honeycomb lattice where the centers of all hexagons are occupied with bosons \cite{Zhangprb}. The effective Hamiltonian for defects is then:
\begin{equation}
\begin{aligned}
H=&-\sum_{i, n, \alpha}\tilde{t}_n^\alpha(a^\dagger_{i}a_{i+\vec{r}(n,\alpha)}+H.C.) +V_{nn}\sum_{\langle i,j\rangle} m_i m_j\\
&+V_{[1,1]}\sum_{\langle \langle i,j \rangle \rangle_{[1,1]}} m_i m_j +V_{[1,-1]}\sum_{\langle \langle i,j \rangle \rangle_{[1,-1]}} m_i m_j \\
&-\tilde{\mu} \sum_i m_i ,
\end{aligned}
\label{effectiveHamiltonian}
\end{equation}
where $\vec{r}(n,\alpha)$ represents the displacement vector corresponding to the hopping term with hopping constant $\tilde{t}^\alpha_n$ and $m_i=a^\dagger_{i}a_{i}$ is the number operator for defects. $\tilde{t}_n^\alpha$ has two sub-indices and $n \in [1,2,3]$, representing the nearest, next-nearest, and next-next-nearest hoppings, respectivley. For $n=1$ or $3$, $\alpha \in [h,d]$ means the hoppings are along horizontal or diagonal direction. While for $n=2$, $\alpha \in [v,d]$ reflects the hoppings in vertical or diagonal direction. Details are shown in Fig. \ref{Fig.6}. We will then expand these hopping terms up to the second order of perturbation, generated from the two-step process of hoppings. Their forms are expressed in the following context. For $n=1$:
\begin{equation}
\begin{aligned}
&\tilde{t}_1^h = \frac{t^2}{V_{[1,-1]}-V_{[1,1]}} \\
&\tilde{t}_1^d = t,
\end{aligned}
\label{effectivet1}
\end{equation}
$n=2$:
\begin{equation}
\begin{aligned}
&\tilde{t}_2^v = \frac{t^2}{V_{[1,-1]}-V_{[1,1]}} \\
&\tilde{t}_2^d = 0,
\end{aligned}
\label{effectivet2}
\end{equation}
and $n=3$:
\begin{equation}
\begin{aligned}
&\tilde{t}_3^h = 0 \\
&\tilde{t}_3^d = \frac{t^2}{V_{[1,-1]}-2V_{[1,1]}}.
\end{aligned}
\label{effectivet3}
\end{equation}
The two-step processes will renormalize the effective chemical potential, leading to $\tilde{\mu}=\mu-2V_{nn}-V_{[1,-1]}+\mu^{(2)}$, where
\begin{equation}
\begin{aligned}
\mu^{(2)} &= \frac{t^2}{V_{nn}-V_{[1,1]}} + \frac{3t^2}{V_{[1,-1]}-V_{[1,1]}} \\
&+\frac{2t^2}{V_{nn}-2V_{[1,1]}}+\frac{2t^2}{V_{[1,-1]}-2V_{[1,1]}} \\
&+\frac{2t^2}{2V_{nn}+V_{[1,-1]}-2V_{[1,1]}} \\
&-\frac{14t^2}{V_{nn}+V_{[1,-1]}-2V_{[1,1]}}.
\end{aligned}
\label{effectivemu}
\end{equation}

Next, we check the stability of the solid by introducing a single defect. The ground state energy, $E_G$, is located at the $\Gamma$ point and for $t>0$: 
\begin{equation}
\begin{aligned}
E_G=&-\tilde{t}_1^h-2\tilde{t}_1^d -2\tilde{t}_2^v -2\tilde{t}_3^d - \tilde{\mu}.
\end{aligned}
\label{EG}
\end{equation}
As $E_G$ changes from positive to negative value, this state with an extra hole/particle on top of the solid will become more stable, which is a signal of phase transition from solid to SS. By substitution, this leads to the curve of $\mu-t$ between solid and SS:
\begin{equation}
\begin{aligned}
\mu=&2V_{nn}+V_{[1,-1]} - \mu^{(2)} -\tilde{t}_1^h-2\tilde{t}_1^d -2\tilde{t}_2^v-2\tilde{t}_3^d.
\end{aligned}
\label{boundary}
\end{equation}
This phase boundary of perturbation theory is shown in Fig. \ref{Fig.7} of the next section, along with our numerical outcomes. 

\subsubsection{\label{sec:level3}CMFT phase diagrams}
\begin{figure*}
\centering
\includegraphics[width=0.9 \textwidth]{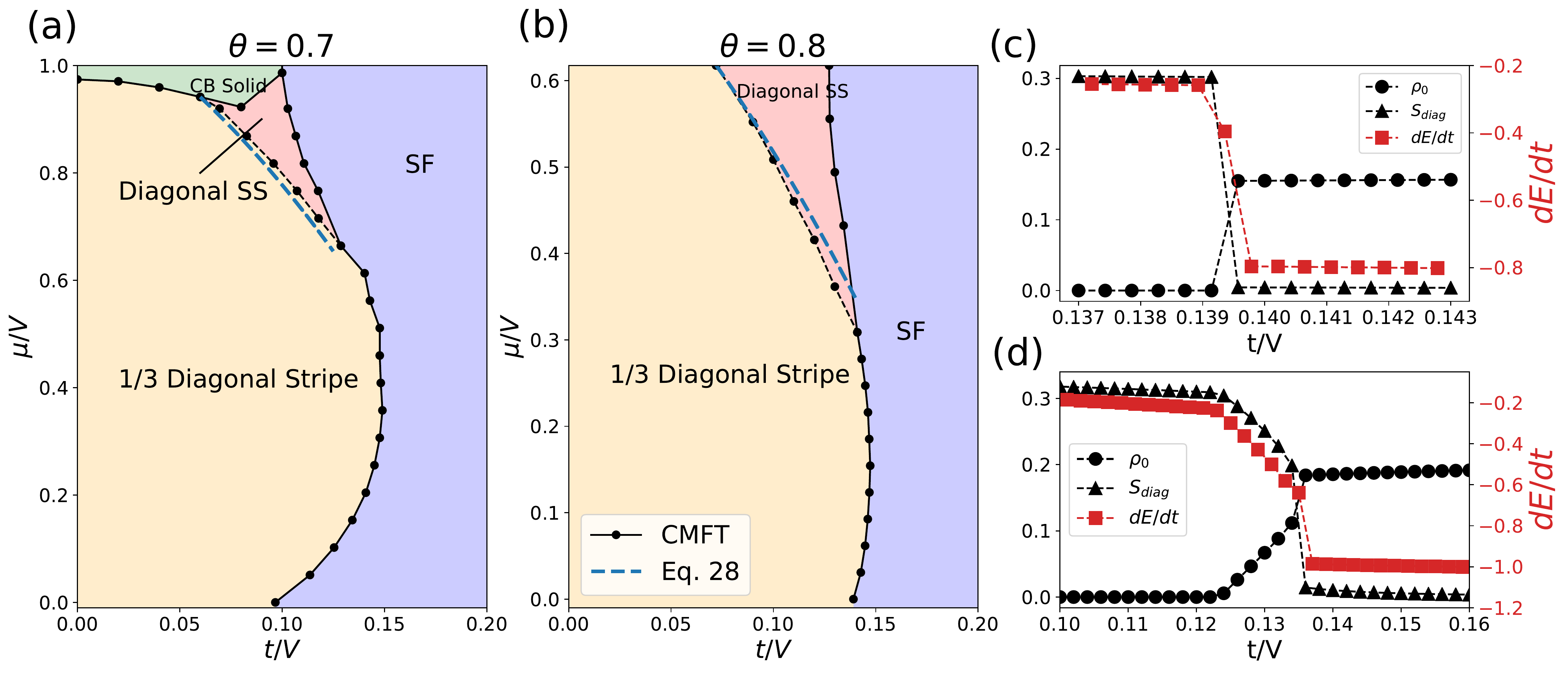}
\caption{\raggedright (a), (b) : Phase diagrams for (a) $\theta = 0.7$ and (b) $\theta = 0.8$ obtained by CMFT. The blue-dashed lines are the analytic phase boundaries between diagonal stripe solid and diagonal SS from the analysis of defect condensation in Eq. 28. (c), (d) : Two horizontal cuts at $\theta = 0.8$ for (c) $\mu = 0$ and (d) $\mu = 0.4$. In (c), a clear first order phase transition between diagonal solid and SF can be seen from the $dE/dt$ curve. In (d), the phase transition between diagonal solid and SS is of second order, while that between SS and SF is of first order.}
\label{Fig.7}
\end{figure*} 
To better determine the phase boundaries, we perform the CMFT calculation to reconstruct the phase diagrams. 
We focus on the scenarios of $\theta = 0.7$ and $\theta = 0.8$, where the diagonal SS phase is more pronounced, as shown in Fig. \ref{Fig.7}(a) and \ref{Fig.7}(b). 
Results are obtained from considering the ground state of $3 \times 3$ and $4 \times 4$ clusters, which can host states that possess a $2\times 2$ unit cell. For the first-order transitions, we determine the phase boundaries by extrapolating the energy on the two sides of the transition to find the intersection. 
On the other hand, the second-order phase boundary between diagonal solid and diagonal SS is defined as the onset of superfluidity, which can be obtained by extrapolating the condensate density on the SS side to zero. The analytic boundaries between solid and SS obtained from defect condensation (Eq. \ref{boundary}) are also shown for comparison.

At small $\mu$, the system at $t = 0$ is in the diagonal stripe phase, characterized by a non-vanishing order $S_{diag} \equiv \left|\tilde{n}(2\pi/3,2\pi/3)+\tilde{n}(2\pi/3,-2\pi/3)\right|$. As $t$ increases, it goes through a first-order phase transition to SF. 
Its first-order nature is characterized by a discontinuous drop in $dE/dt$, as can be seen in Fig. \ref{Fig.7}(c), where a horizontal cut along $\mu = 0$ for Fig. \ref{Fig.7}(b) is shown. 

As $\mu$ becomes larger, a diagonal SS phase appears between the solid and SF phase, where both $S_{diag}$ and $\rho_0$ are finite. 
As an example, the cut along $\mu=0.4$ is shown in Fig. \ref{Fig.7}(d). 
Focusing on the $dE/dt$ curve, we observed a discontinuous drop at the transition point from SS to SF, indicating a first-order phase transition. 
This is consistent with the aforementioned mean-field result. 
The discontinuity reduces as $\mu$ increases and disappears at the symmetry point, $\mu = 2V_{nn}+V_{[1,1]}+V_{[1,-1]}$, where the transition becomes second-order. 
We do not observe such drop in $dE/dt$ at the onset of superfluidity from the diagonal stripe phase, which suggests that the transition from solid to SS is of second order. 
In Fig. \ref{Fig.7}(a) and (b), the phase boundaries between solid and SS, obtained by defect condensation, are plotted as blue dashed curves, showing good agreement with the CMFT counterparts. 
It is worth noting that these two boundaries agree better at higher $\mu$. This is because the boundary is very shallow in the diagonal SS phase and about to change into the SF phase at lower $\mu$. Since the quantum effect is the strongest in the SF phase, where the hopping term in the Hamiltonian becomes dominant, our two methods, the perturbation and CMFT, lose their effectiveness because they both neglect part of the quantum fluctuation.
It suggests that the perturbation theory performs well in this transition.
At $\theta = 0.7$, a CB solid phase appears as the chemical potential approaches the symmetry point. 
Since the CB solid breaks different translational symmetry, the phase boundary is also first-order. This again confirms the competing picture that we have revealed by mean-field analysis.

Comparing Fig. \ref{Fig.7}(a) with its corresponding mean-field phase diagram in Fig. \ref{Fig.4}(d), different phase boundaries are quite consistent except for the one between SS and SF, where the mean-field theory overestimates the SS regions. 
This is due to the fact that mean-field approaches ignore the effect of quantum fluctuation, which is unfavorable for long-range orders. In our model, quantum fluctuation becomes the strongest when the effective external field $h$ approaches zero. Therefore, the mean-field phase boundary is least accurate in this regime.

\subsubsection{\label{sec:level3}Thermodynamic limit}

\begin{figure}
\centering
\includegraphics[width=0.48 \textwidth]{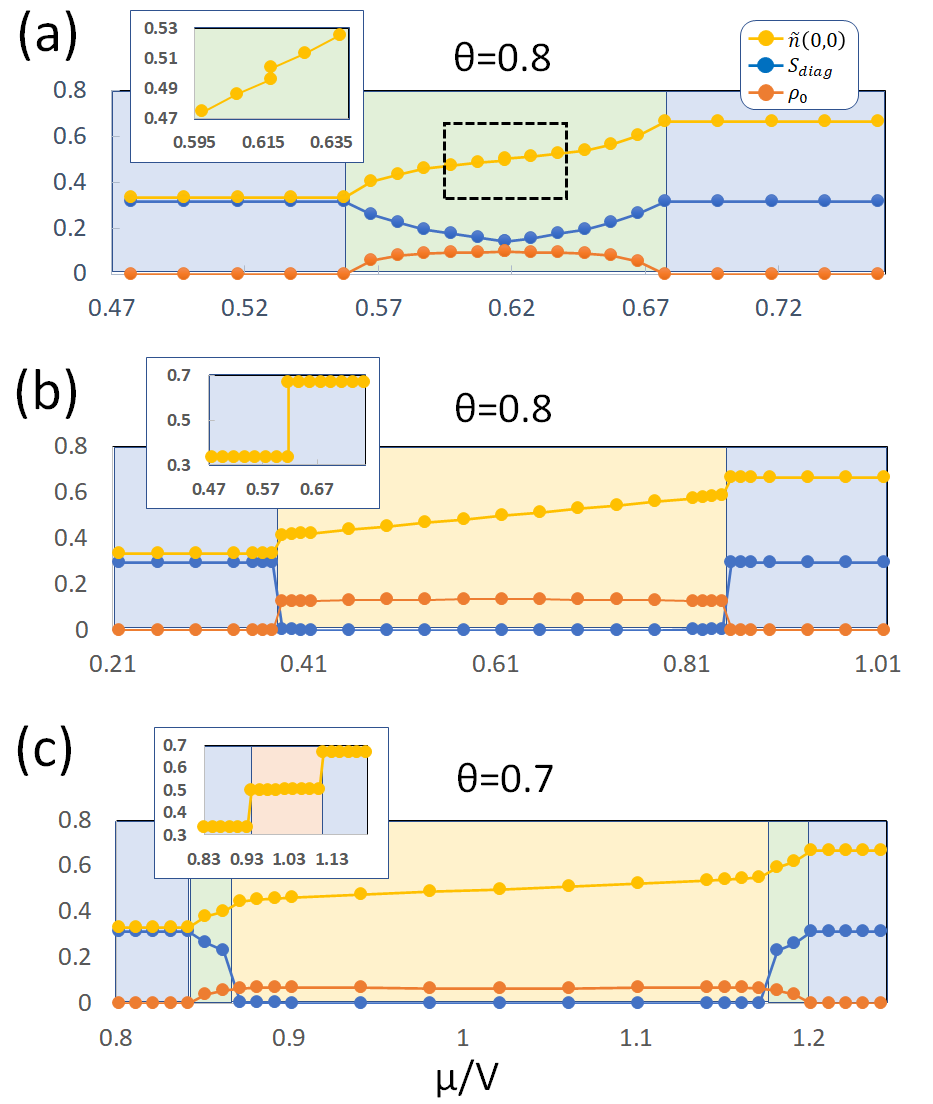}
\caption{\raggedright iPEPS data for order parameters $\tilde{n}(0,0)$, $S_{diag}$, and $\rho_0$. In (a), the cut is along $t/V=0.085$ for $\theta=0.8$. Green area marks the region for SS phase while solid phases are indicated with blue background. The inset magnifies the region enclosed by black dotted box in the middle. In (b), the cut is now along $t/V=0.12$. For this cut, SS is replaced by SF denoted by yellow background. Its inset shows a cut along $t/V=0.06$. For $\theta=0.7$, a cut along $t/V=0.085$ is present in (c). The inset demonstrates again a cut for $t/V=0.06$, where competition between CB and diagonal stripe solids takes place.}
\label{Fig.8}
\end{figure} 

An important point we need to look into is whether the above phases obtained by mean-field-based methods can be stable after being pushed to the thermodynamic limit. 
Of course, this can be done by gradually increasing the cluster size in the CMFT and performing the scaling technique, which will be shown in the Appendix.
But here, we apply another strategy and use iPEPS for this purpose.
However, we have noticed that the simple-update iPEPS is no longer enough for the reason that these phases possess competing energies very close to each other and it is hard to distinguish them without a better optimization algorithm. 
Therefore, we turn to another kind of iPEPS, which is based on variational optimization \cite{LiaoS, Hasik}.
For this kind of iPEPS, we first encode the full progress when obtaining the variational energy of the target Hamiltonian, starting from the wavefunctions in the form of tensors. 
We then apply the backward-propagated automatic differentiation (AD) to calculate the gradients of our target function, the variational energy, and then optimize the wavefunctions.
In this way, we do not have to go through the traditional process of optimization where estimation after singular value decomposition (SVD) happens repeatedly. 

Fig. \ref{Fig.8} represents the results of our variational iPEPS with bond dimension $D=4$. 
In Fig. \ref{Fig.8}(a), we go through a vertical cut at $t/V=0.085$ in Fig. \ref{Fig.7}(b). 
In the middle we can clearly see a region where condensate density ($\rho_0$) coexists with solid order $S_{diag}$, indicating the existence of diagonal SS. 
We enlarge the region encircled by the black dotted box and it is clear that at $\tilde{n}(0,0) \approx 0.5$ there is a two-fold degeneracy. 
This is because our two diagonal stripe solids are analogous to the Ising spins under an external magnetic field. 
When $\mu$ is larger than the symmetric point (effective field pointing upward), the state prefers the one with more bosons (spin aligning upward), and $\mu$ being smaller than the symmetric point (effective field pointing downward) is the other way around. 
Such an effect can be seen more clearly in the inset of Fig. \ref{Fig.8}(b), where the cut along $t/V=0.06$ is drawn. 
Due to the above-mentioned reason, the quantum state will resume half-filled only when the solid order is smeared out; that is to say, when entering the SF phase.
As a result, SS can not be half-filled in the current scope since its solid order is not zero. 
This conclusion is the same as that in Ref. \onlinecite{Zhangnjp}.

Fig. \ref{Fig.8}(b) shows another vertical cut along $t/V=0.12$. 
An obvious first-order phase transition can be seen between diagonal solid and superfluid, indicated by a sudden jump of order parameters. 
To reveal the competition between phases of different sublattices, we plot the cut of $t/V=0.085$ at $\theta=0.7$. 
Along this cut, diagonal stripe solid melts into diagonal SS, right before the solid order is completely smeared out with a first-order transition into the superfluid. 
However, unlike the CMFT phase diagram, we do not see the phase competition here.
We then plot another cut along $t/V=0.06$ in the inset for showing the competition between different solid phases. 
It is clear that diagonal stripe solid transits into CB solid with a first-order phase transition. 
Here, we apply the variational iPEPS with different unit cells, in order to obtain both states. 
We then compare their energies to determine the ground state, as we have done in the mean-field analysis.

We have noticed that some differences appear in comparison with the CMFT phase diagrams. 
For example, in Fig. \ref{Fig.8}(c) the SS phase goes through a first-order phase transition into the superfluid. 
However, in Fig. \ref{Fig.7}(a), SS would directly transit into the checkerboard solid in the same cut. 
This indicates the overestimation of SS phase from CMFT, which is not surprising because mean-field treatment has this tendency. 
Recall that we mentioned that CMFT can capture the short-range correlation within the cluster, but for SF it contains the off-diagonal long-range order (ODLRO). 
Therefore, within a finite cluster such long-range effect is underestimated, while the solid order is properly described if the cluster is larger than its unit cell.

Among all of these, a more detailed investigation from iPEPS, such as enlarging the bond dimension, may be able to provide a more precise phase diagram much closer to the real experiments. 
We will, nevertheless, leave this part to future works since in this report we mainly focus on elucidating the competition between phases and their identifications.  

\section{\label{sec:level1}Discussion}

We have considered the short-range dipolar model with tilting polar angles. 
For the polarization lying in the x-z(y-z) plane, we have re-constructed the similar phase diagram as that shown in Ref. \cite{Zhangnjp} with CMFT and simple-update iPEPS.
By setting the dipole moment pointing along the direction with $\phi=\pi/4$ while varying the polar angle, we have discovered that the physical scenario is in fact the competition among phases possessing different sizes of the unit cell. 
Moreover, the diagonal stripe and its SS correspond to quantum states in an effective triangular lattice. 
Thus, our results reveal the possibility to generate a scenario for a triangular optical lattice out of the original square lattice, with the fine tuning of dipolar angle.

Previously, we have mentioned that the difference between our model and the long-range dipolar model is that the so-called Devil's staircase is excluded in the short-range model. 
In addition to this, it is common for some SS phases to become destabilized for the short-range model due to the strong quantum fluctuation \cite{Yamamoto3}.
Nevertheless, our main phases in Section II.C are shown to be present at the thermodynamic limit. Therefore, it is reasonable to expect that our SS phases can also exist for the full long-range dipolar model.
Moreover, we have noticed that for isotropic dipolar interaction, the diagonal stripe solid can already be seen \cite{Capogrosso}.
Therefore, for the long-range model, the diagonal stripe could become the prominent phase at smaller $\theta$ than our prediction.

For the scenario of $\phi = \pi/4$ considered in this work, the projection of polarization onto the XY plane is along the $[1,1]$ direction. 
Therefore, as we tilt the polarization along this direction, its repulsive potential becomes smaller and eventually turns into attractive interaction, leading to the formation of diagonal stripe.
If we tune $\phi$ away from $\pi/4$, since the energy for diagonal order is less favorable, we expect that the regime of competing CB phase will become larger in the phase diagram. 
For the other limit, $\phi = 0$, the projection is now along the x direction, giving us the stripe order at large $\theta$, as shown in Fig. \ref{Fig.2}(a). 
Therefore, we expect there to be a critical $\phi_c$ where the crossover from diagonal stripe to normal stripe takes place at large $\theta$. 
A quick estimation for $\phi_c$ will be when the nnn interaction in the [1,1] direction is equal to the nn interaction in the x-direction.
Moreover, for the supersolid phase we expect to see a mixture of normal and diagonal stripe for intermediate $\phi$, because of an extra anisotropy now for the nn interaction. 
Although we still expect its unit cell to be of $3\times3$ because the interaction is still only composed of nn and nnn interactions, a three-sublattice description in Fig. \ref{Fig.3}(b) may not be enough because of the mixture with normal stripe order.
But since we do not expect any new phase except the stripe solid and its SS with the current Hamiltonian (Eq. \ref{Hamiltonian1/4phi}), we did not examine other $\phi$ values here. 
We believe a more interesting scenario would be gradually increasing the interaction range, where an interpolation to the long-range physics can be realized and yet avoid the hard-to-track Devil's staircase \cite{Masella}. 
Of course, a true long-range model is believed to reflect the underlying physics to the highest degree.
We realize that for such a long-range interactive model, QMC would be better for numerical calculation. 
QMC is known for its capability of providing numerically exact solutions for many-body Hamiltonians \cite{Berg}. 
However, the notorious sign problem often takes place for fermionic systems or Hamiltonians with frustration \cite{Loh}, and therefore hinders the usage of QMC. 
Luckily the long-range dipolar Hamiltonian that we are going to consider does not suffer from such issues and is expected to be numerically solved with QMC.
Therefore, a more detailed research considering the long-range interaction with the help of QMC, is left for future consideration.

Finally, we would like to discuss the experimental realization of these different phases. Note that, there are several energy scales, that is, the onsite Hubbard interaction $U$, the dipolar interaction strength $V$, the hopping amplitude $t$, and the thermal energy $k_BT$. The regime of interest in our phase diagram is around $V/t \eqsim 10$. Together with the hard-core requirement, the desired energy relation is
\begin{align}
	U \gg V \approx 10t 
	\label{eq:energy_scales}
\end{align}
In an optical lattice, $U$ and $t$  are determined by the recoil energy $E_r$ and the depth of the lattice potential $V_0$. 
for the hard-core limit, one needs the fraction $s = V_0/E_r$ to be greater than 20 \cite{jaksch1998cold}, which can be achieved by tuning the laser intensity. 
While $t$ and $U$ can be varied more easily, the most challenging part for the experiments appears in the relation between $V$ and $k_BT$. 
Although we did not perform the finite temperature calculation, the detection of superfluidity in SS and SF usually requires the thermal energy scale to be lower than the kinetic energy ($k_BT<t$). 
Together with Eq. \ref{eq:energy_scales}, we will need $V > 10 k_BT$.
Taking the Feshbach molecule $\mathrm{Er_2}$ \cite{frisch2015ultracold} as an example, whose dipole moment is $14\mu_B$, the nearest-neighbor dipolar interaction is $V/\hbar \approx 3.18\times 10^2 \mathrm{hz}$ for lattice constant to be 400 nanometers; this would require the temperature to be lower than $1 \mathrm{nK}$. 
Another possibility would be the electrically dipolar molecules \cite{gadway2016strongly}. 
In this case, the natural interactive strength can be larger than that of the magnetic atoms by several orders of magnitude. However, the problem of decoherence from sources caused by photon-induced scattering and chemical reactions \cite{quemener2012ultracold,mayle2013scattering,guo2018dipolar,christianen2019photoinduced,hu2019direct} still remains a challenge for such systems. Nevertheless, thanks to the advance of experimental techniques, we believe the realization of various phases presented in this work is definitely possible in the near future.

\section{\label{sec:level1}Acknowledgement}
W.-L.T. would like to thank Juraj Hasik, Naoki Kawashima, and Tsuyoshi Okubo for the decent discussions.
H.-K.W. is supported by JQI-NSF-PFC (supported by NSF grant PHY-1607611). 
W.-L.T. is supported by Postdoctoral Research Abroad Program, Project No. 108-2917-I-564-007, from Ministry of Science and Technology (MOST) of Taiwan. 
%





\appendix

\renewcommand\thefigure{\thesection.\arabic{figure}}
\setcounter{figure}{0}

\section{Scaling analysis of CMFT}

\begin{figure}
\centering
\includegraphics[width=0.48 \textwidth]{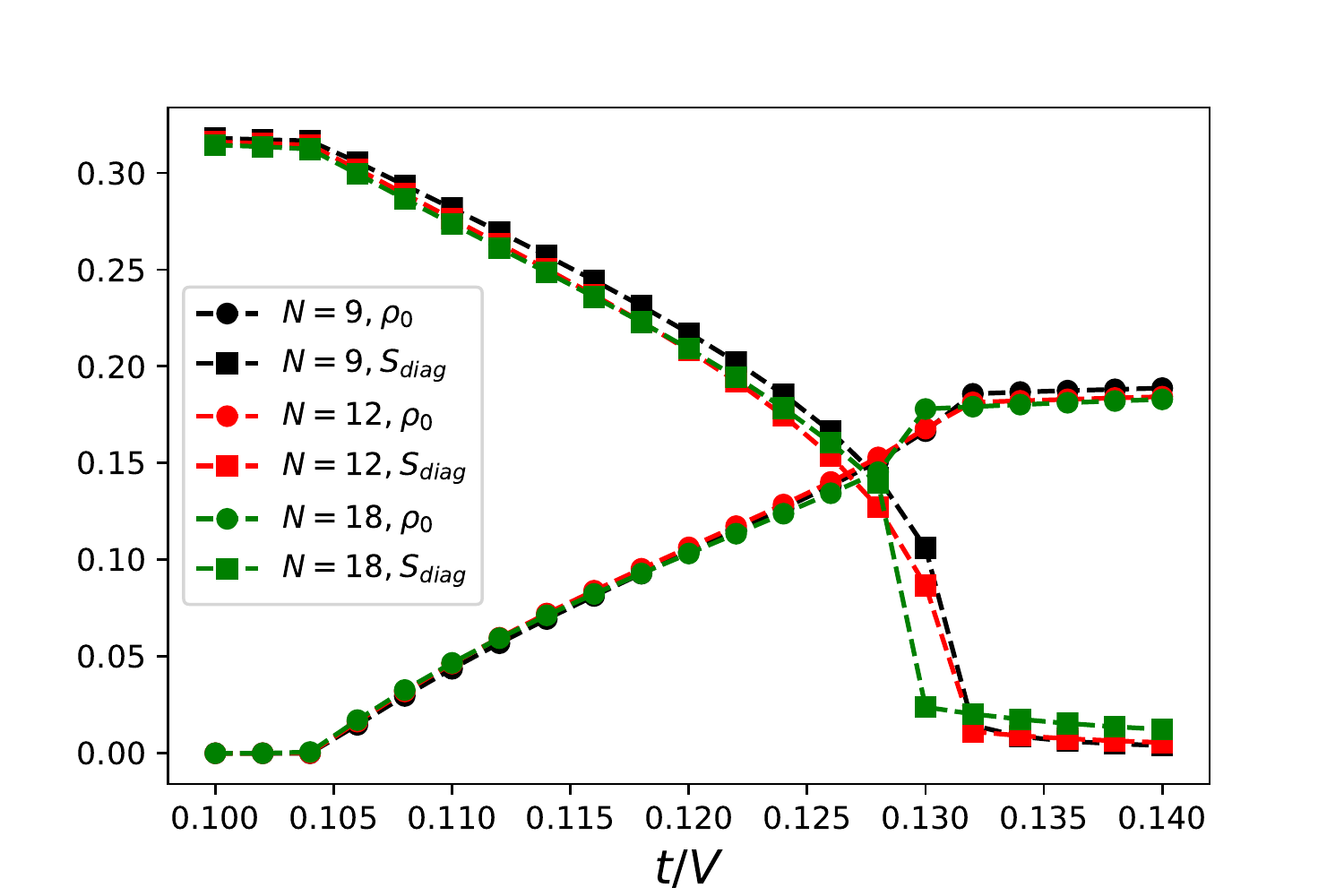}
\caption{\raggedright Order parameters for different cluster along line cut $\mu = 0.494$ at $\theta = 0.8$.}
\label{Fig.A1}
\end{figure} 
\begin{figure}
\centering
\includegraphics[width=0.48 \textwidth]{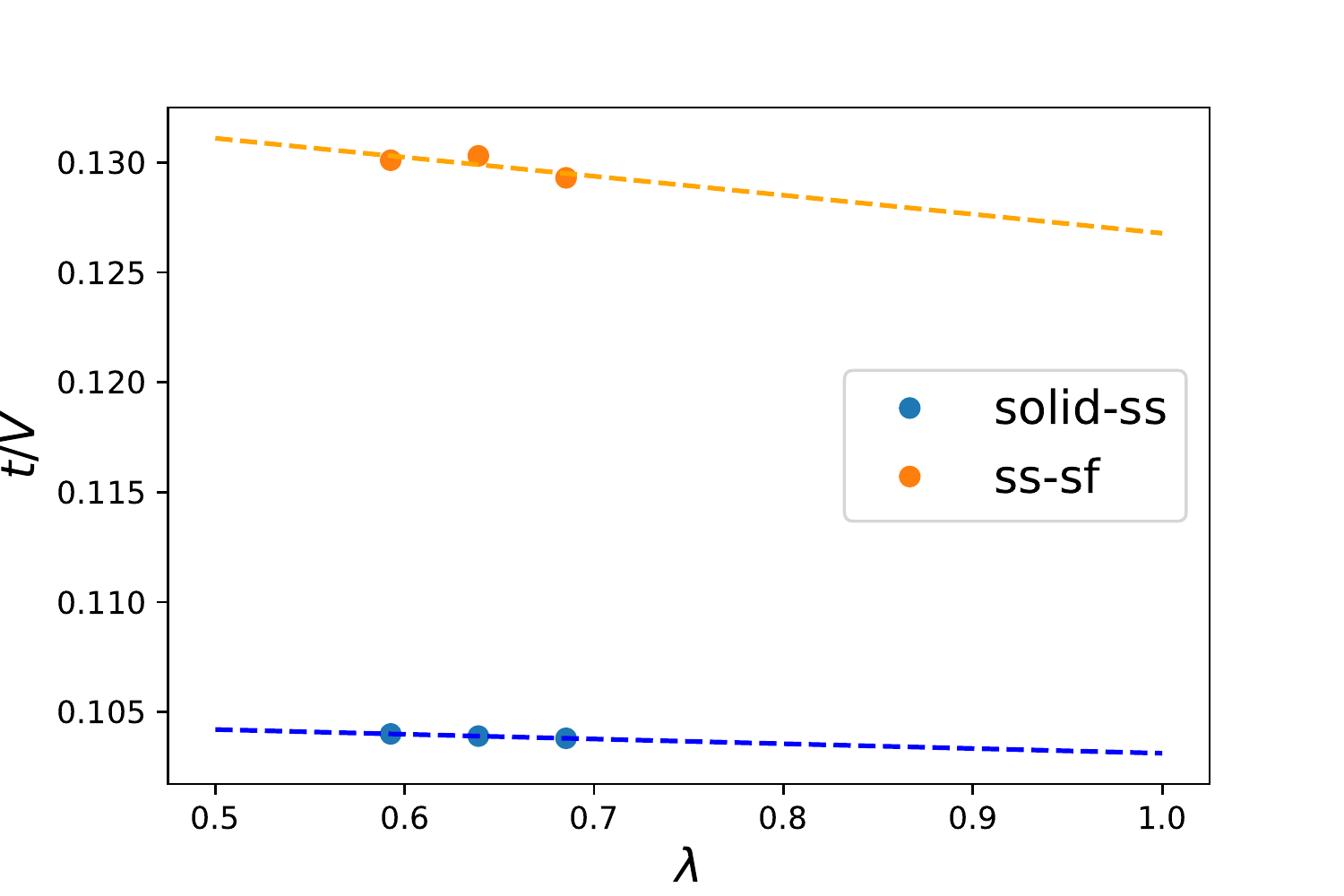}
\caption{\raggedright Scaling analysis for the phase boundaries in Fig. \ref{Fig.A1}. The scaling parameter $\lambda$ is defined using the effective triangular lattice. }
\label{Fig.A2}
\end{figure} 
In CMFT calculations, the effect of correlation with range beyond the cluster size is excluded. 
To infer the physics in the thermodynamic limit, a common method is to repeat the calculation on clusters of different sizes and perform extrapolation. 
We employ the scaling method introduced in Refs. \onlinecite{Yamamoto2,Yamamoto3}, with the scaling parameter $\lambda$ defined as $\lambda\equiv N_B/(N_C\times z/2)$, where $N_B$ is the number of nn bonds, $N_C$ is the number of sites, and $z$ is the coordination number for the lattice. 
Since $\lambda \rightarrow 1$ in the thermodynamic limit, we can approach this limit by extrapolating $\lambda$ to $1$.

Here we examine the phase boundary for diagonal SS as $\lambda\rightarrow 1$. 
We choose the line cut along $\mu = 0.494$ at $\theta = 0.8$, where the diagonal SS phase can be found between $0.104<t/V<0.13$ in the 3$\times$3 cluster, as shown in Fig. \ref{Fig.7}(b). Besides the 9-site cluster, we have conducted the calculation with 12- and 18-site clusters. 
The 12-site cluster is a 3 $\times$ 4 rectangular lattice and the 18-site cluster is a $45^{\circ}$-tilted square lattice defined by the side vectors $(3,3)$ and $(3,-3)$. 
Results are shown in Fig. ~\ref{Fig.A1}, where the order parameters from these three clusters are consistent with each other, except when approaching the SS-SF transition point. We notice that SS phase becomes thinner as cluster size increases, which accords with the observation by iPEPS.

The finite-size scaling is performed for the phase boundaries. Since the phases appearing here are from the equivalent triangular lattice, we instead adopt the lattice structure in Fig. \ref{Fig.3}(c) to define the scaling parameter $\lambda$. The result is shown in Fig. \ref{Fig.A2}. According to the linear extrapolation, at $\lambda = 1$ the ss-sf(solid-ss) transition point is at $t/V \sim 0.1031(0.1268)$. These values are reasonably close to those from the 3$\times$3 cluster, which are $0.1040$ and $0.1301$. Therefore, we can conclude that the 3$\times$3 cluster already provides a good estimation for the CMFT.

\renewcommand\thefigure{\thesection.\arabic{figure}}
\setcounter{figure}{0}
\section{iPEPS phase boundary}
\begin{figure}
\centering
\includegraphics[width=0.48 \textwidth]{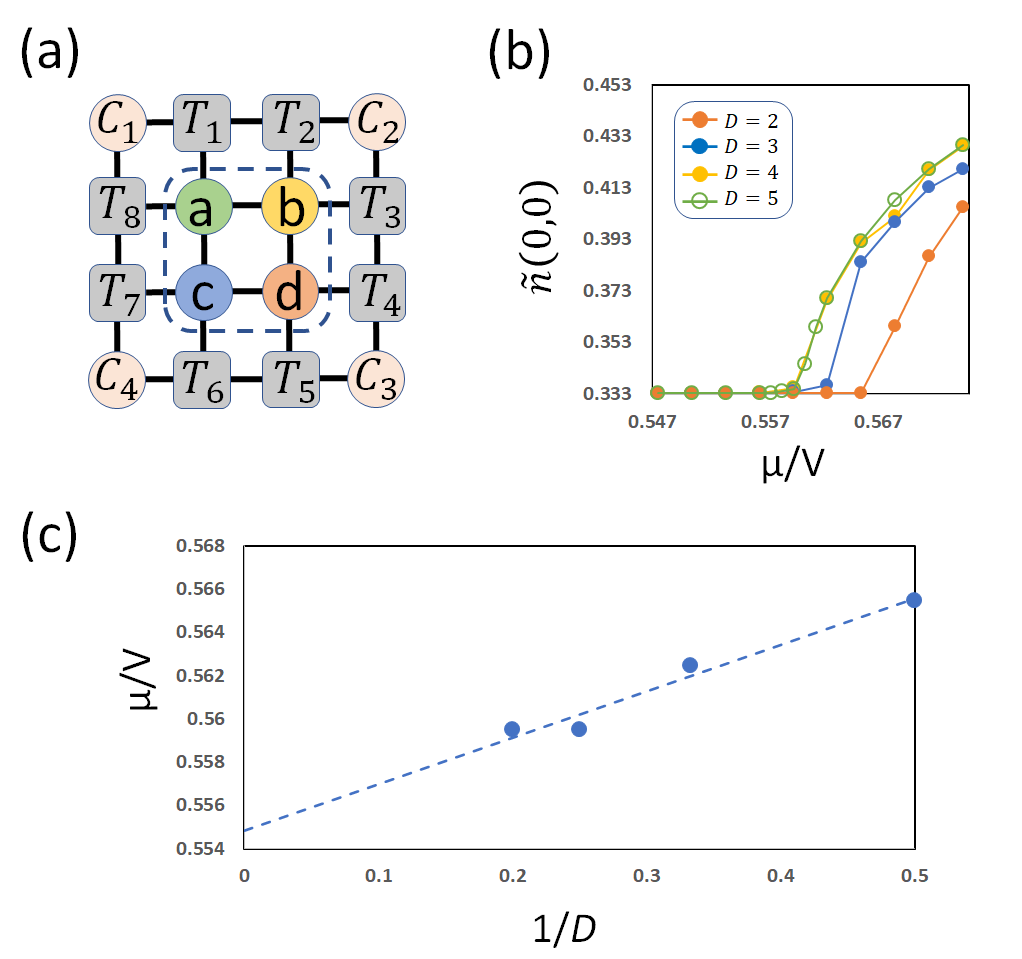}
\caption{\raggedright (a) The demonstration of a $2\times2$ unit cell with its environment. For this choice of unit cell, we have four corner tensors, $C_1-C_4$, and eight edge tensors, $T_1-T_8$. (b) The plot of average density ($\tilde{n}(0,0)$) for different choices of bond dimension $D$, with $\theta=0.8$ and $t/V=0.085$. It is clear that the curves for $D=4$ and $D=5$ are nearly overlapping. (c) The scaling analysis for the transition point along with $1/D$. The extrapolated transition point is $\mu/V=0.5549$.}
\label{Fig.B1}
\end{figure} 
As mentioned in the previous content, we numerically attain the thermodynamic limit with the help of iPEPS. 
This tensor network ansatz is made of two parts. 
The first part is the bulk tensors, whose number is determined by the unit cell we choose. 
Fig. \ref{Fig.B1}(a) demonstrates a $2\times2$ unit cell with four bulk tensors, after contracting the physical index of the original rank-5 tensors and forming the so-called double-layered tensors.
These tensors are the targets for the optimization, with the methods of imaginary-time evolution \cite{JiangS, Jordan} or variation \cite{LiaoS, Hasik}. 
After the bulk tensors being optimized, we will then construct the environment tensors, made of edge ($T$) and corner ($C$) tensors. 
The corner tensors have two indices of dimension $\chi$, while edge tensors have another additional index of dimension $D^2$, which is connected with the bulk tensors. 
We follow the way of corner-transfer-matrix renormalization group (CTMRG) in Ref. \onlinecite{Corboz} and obtain our environment, which is the effective extrapolation to the infinite size. 
We then use the optimized bulk tensors and environment to evaluate physical properties such as energy or observables.

In order to verify our choice of bond dimension $D=4$ in the main text, here we compare the results by various $D$ and apply the scaling analysis for the transition point.
In \ref{Fig.B1}(b), we demonstrate the plot of average density ($\tilde{n}(0,0)$) to chemical potential for different choices of $D$, with $t/V=0.085$ and $\theta=0.8$.
It is clear that for $D\ge4$, curves overlap nearly perfectly. 
This outcome indicates that for our current model, $D=4$ is already a fairly good choice.
Therefore, we have reason to believe that our results of iPEPS in the main text are insensitive to the further enlargement of $D$.
Fig. \ref{Fig.B1}(c) demonstrates a scaling analysis of the transition point by different $D$.
The ultimate transition point after extrapolation is $\mu/V=0.5549$, which is very close to the transition point for $D=4$.

\bibliography{draft}

\end{document}